\documentclass[12pt]{iopart}

\usepackage{graphicx,setspace}
\usepackage{dsfont}
\usepackage{amssymb}
\usepackage{mathrsfs}
\usepackage{psfrag}

\begin{document}
\title{Canonical quantization of macroscopic electromagnetism}
\author{T G Philbin}
\address{School of Physics and Astronomy, University of St Andrews,
North Haugh, St Andrews, Fife KY16 9SS,
Scotland, UK.}
\ead{tgp3@st-andrews.ac.uk}

\begin{abstract}
Application of the standard canonical quantization rules of quantum field theory to macroscopic electromagnetism has encountered obstacles due to material dispersion and absorption. This has led to a phenomenological approach to macroscopic quantum electrodynamics where no canonical formulation is attempted. In this paper macroscopic electromagnetism is canonically quantized. The results apply to any linear, inhomogeneous, magnetodielectric medium with dielectric functions that obey the Kramers-Kronig relations. The prescriptions of the phenomenological approach are derived from the canonical theory.
\end{abstract}

\pacs{42.50.Nn, 12.20.-m}

\section{Introduction}
Macroscopic electromagnetism is an indispensable part of theoretical physics, providing an accurate description of a large variety of light-matter interactions~\cite{LLcm,jac}. The ability to encompass the electric and magnetic responses of media within macroscopic dielectric functions, without reference to the microscopic material structure, is of great practical importance. A microscopically complete treatment of the interaction of light with most real materials is currently impossible, and even if the computer resources necessary for such a task were available, simple analytical calculations using the macroscopic theory would still be adequate in many circumstances of interest.  

But the source of the strength and versatility of macroscopic electromagnetism, namely the removal of microscopic material degrees of freedom, has been seen as a disadvantage when it comes to quantizing the theory~\cite{hil84,hut92}. A rigorous quantization procedure would require a Lagrangian and Hamiltonian formulation of the theory, followed by the standard canonical quantization rules. Such a canonical formulation of macroscopic electromagnetism has so far been thwarted by the presence in all media of dispersion and absorption, an obstacle that cannot be avoided if one considers all frequencies of electromagnetic waves for which the macroscopic theory is valid. This difficulty has led to a phenomenological approach to macroscopic quantum electrodynamics (QED), in which no canonical formulation is attempted (some details of this scheme will be described below; for a comprehensive description see~\cite{kno01,sch08} and references therein). The lack of a canonical basis for macroscopic QED is ameliorated by the results of calculations that introduce a simple microscopic model for a dielectric~\cite{hut92,sut04a,sut04b,khe08}. Huttner and Barnett~\cite{hut92} showed that the canonical quantization of the resulting coupled light-matter system leads to results that can be expressed in terms of a homogeneous electric permittivity, derived from the matter degrees of freedom, at which point the model reproduces relations introduced in macroscopic QED on a phenomenological basis~\cite{kno01,sch08}. The simple microscopic model has been extended to the case of an inhomogeneous dielectric~\cite{sut04a,sut04b} and to a magnetodielectric medium~\cite{khe08}, again producing results at the level of the dielectric functions that agree with the prescriptions of the phenomenological theory~\cite{kno01,sch08}. An essential ingredient of the microscopic models is the introduction of a reservoir in addition to the electromagnetic and matter degrees of freedom, in order to incorporate the dissipation that is necessarily present due to the Kramers-Kronig relations.

The drawback of the microscopic models as a justification of the phenomenological approach is, of course, their simplicity. As a classical theory, macroscopic electromagnetism is experimentally proven to be valid in the case of highly complicated materials, and a rigorous quantization would ideally also proceed on a macroscopic level. This requires a canonical formulation and quantization of macroscopic electromagnetism, a task that is performed here. The resulting theory is valid for all linear, inhomogeneous, magnetodielectric media with dielectric functions that obey the Kramers-Kronig relations. The treatment is macroscopic, in that the electromagnetic fields are shown to obey the macroscopic Maxwell equations for an arbitrary linear medium, but a reservoir is necessary to account for dissipation, as in the microscopic models. Indeed, it is clear from the outset that a reservoir must be included in addition to the electromagnetic fields, since otherwise one would obtain a Hamiltonian involving only the latter, and thereby a conserved electromagnetic energy in dissipative media. But the necessity of including a reservoir does not imply that additional dynamical degrees of freedom for the medium must also be introduced; only the electromagnetic fields and the reservoir are required.  This last point has been recognized in some previous treatments of QED in media~\cite{bha06,khe06,sut07,amo08,amo09,khe10}; we here give a comprehensive derivation of general, linear macroscopic QED from electromagnetic fields coupled to a reservoir. Due to the common feature of a reservoir, the details of the canonical macroscopic theory have much in common with the  microscopic models~\cite{hut92,sut04a,sut04b,khe08}. In fact, it is no more difficult to canonically quantize macroscopic electromagnetism than to quantize the simple microscopic models. But the canonical theory is a greater reward for a similar amount of labour, as it is applicable to any medium where the classical macroscopic theory is valid. 

The postulates of the phenomenological theory~\cite{kno01,sch08} that serve as its starting point are here derived from canonical quantization. We thereby remove the need for a phenomenological approach since the canonical theory has the same range of validity (in contrast to the microscopic models). As well as providing a foundation for the wide range of important results that have been obtained from the phenomenological prescriptions~\cite{kno01,sch08}, the canonical theory will allow investigation of issues that require a Hamiltonian and the related canonical machinery to be properly addressed. An important example where the benefits of a canonical theory are desirable is the Casimir-Lifshitz effect~\cite{cas48,lif55,dzy61,LL}, the phenomenon of forces on bodies due to electromagnetic zero-point and thermal fields. The standard Lifshitz theory~\cite{lif55,dzy61,LL} of this effect, being macroscopic and lacking a canonical foundation, suffers from the drawbacks of the phenomenological approach to macroscopic QED in general. This has led to doubts about the status of Lifshitz theory as a proper quantum treatment; the lack of an intelligible Hamiltonian has been criticized\footnote{I am indebted to Gabriel Barton for properly stressing this deficiency of Lifshitz theory.} and it has recently been argued that the theory is essentially classical~\cite{ros10}. The canonical macroscopic QED developed here can provide a rigorous foundation for Lifshitz theory.

The action principle from which all the results follow is written in Section~\ref{sec:action} and shown to lead to the macroscopic Maxwell equations. In Section~\ref{sec:quantization} the standard rules of canonical quantization are applied and the Hamiltonian is derived. The Hamiltonian is diagonalized in Section~\ref{sec:diag} and the electromagnetic field operators are written in terms of the diagonalizing operators; at this point the postulates of the phenomenological approach emerge.

\section{The action and the field equations} \label{sec:action}
In the canonical formulation of electromagnetism~\cite{wei} the dynamical fields must be taken to be the scalar potential $\phi$ and vector potential $\mathbf{A}$, related to the electric and magnetic fields by
\begin{equation} \label{pots}
\mathbf{E}=-\nabla\phi-\partial_t\mathbf{A}, \qquad \mathbf{B}=\nabla\times\mathbf{A}.
\end{equation}
The relationship between fields in the time and frequency domains is, for the example of the electric field,
\begin{equation}  \label{Efreq}
\mathbf{E}(\mathbf{r},t)=\frac{1}{2\pi}\int_0^\infty \rmd\omega\left[\mathbf{E} (\mathbf{r},\omega)\exp(-\rmi\omega t)+\mbox{c.c.}\right].
\end{equation}
In limited frequency ranges where losses are negligible the action for dispersive macroscopic electromagnetism can be written as a functional of $\phi$ and $\mathbf{A}$ only~\cite{phi10}. Our concern here, however, is with the theory in its full generality, where all frequencies are included and the dielectric functions obey the Kramers-Kronig relations~\cite{jac,LLcm}. This means that in addition to $\phi$ and $\mathbf{A}$ the Lagrangian must include dynamical variables that account for absorption and dissipation of the electromagnetic energy. We incorporate dissipation in exactly the manner used in the microscopic models~\cite{hut92,sut04a,sut04b}, by introducing a continuum of reservoir oscillator fields. As we allow for a magnetic as well as an electric response in the macroscopic medium, the reservoir will be a set of two such continua of oscillator fields, which we denote by $\mathbf{X}_\omega(\mathbf{r},t)$ and $\mathbf{Y}_\omega (\mathbf{r},t)$, where $\omega>0$ is a continuous label giving the frequency of oscillation. The dynamical variables are thus $\phi$, $\mathbf{A}$, $\mathbf{X}_\omega$ and $\mathbf{Y}_\omega$, and we must write an action that is a functional only of these fields and whose dynamical equations are the macroscopic Maxwell equations in a general inhomogeneous magnetodielectric medium. As there are no dynamical variables representing the medium, it must be included only through the coupling of the electromagnetic fields with the reservoir. 

The required action is
\begin{equation} \label{S}
\fl
S[\phi,\mathbf{A},\mathbf{X}_\omega,\mathbf{Y}_\omega]=S_{\mathrm{em}}[\phi,\mathbf{A}]+S_ \mathrm {X}[\mathbf{X}_\omega]+S_ \mathrm{Y}[\mathbf{Y}_\omega]+S_{\mathrm{int}}[\phi,\mathbf{A},\mathbf{X}_\omega,\mathbf{Y}_\omega],
\end{equation}
where $S_{\mathrm{em}}$ is the free electromagnetic action:
\begin{equation} \label{Sem}
S_{\mathrm{em}}[\phi,\mathbf{A}]=\frac{\kappa_0}{2}\int\rmd^4 x\left(\frac{1}{c^2}\mathbf{E}\cdot\mathbf{E}-\mathbf{B}\cdot\mathbf{B}\right), \quad \kappa_0=1/\mu_0,
\end{equation}
$S_ \mathrm{X}$ and $S_ \mathrm{Y}$ are the actions for the free reservoir oscillators:
\begin{eqnarray}
S_ \mathrm{X}[\mathbf{X}_\omega]=\frac{1}{2}\int\rmd^4 x\int_0^\infty\rmd\omega\left(\partial_t\mathbf{X}_\omega\cdot\partial_t\mathbf{X}_\omega-\omega^2\mathbf{X}_\omega\cdot\mathbf{X}_\omega\right),  \\[3pt]
S_ \mathrm{Y}[\mathbf{Y}_\omega]=\frac{1}{2}\int\rmd^4 x\int_0^\infty\rmd\omega\left(\partial_t\mathbf{Y}_\omega\cdot\partial_t\mathbf{Y}_\omega-\omega^2\mathbf{Y}_\omega\cdot\mathbf{Y}_\omega\right),
\end{eqnarray}
and $S_{\mathrm{int}}$ is the interaction part of the action, coupling the electromagnetic fields to the reservoir:
\begin{eqnarray}
S_{\mathrm{int}}[\phi,\mathbf{A},\mathbf{X}_\omega,\mathbf{Y}_\omega]=\int\rmd^4 x\int_0^\infty\rmd\omega\left[\alpha(\mathbf{r},\omega)\mathbf{X}_\omega\cdot\mathbf{E}+\beta(\mathbf{r},\omega)\mathbf{Y}_\omega\cdot\mathbf{B}\right],  \label{Sint} \\[3pt]
\alpha(\mathbf{r},\omega)=\left[\frac{2\varepsilon_0}{\pi}\omega\varepsilon_\mathrm{I}(\mathbf{r},\omega)\right]^{1/2}, \qquad \beta(\mathbf{r},\omega)=\left[-\frac{2\kappa_0}{\pi}\omega\kappa_\mathrm{I}(\mathbf{r},\omega)\right]^{1/2}.  \label{ab}
\end{eqnarray}
In the coupling functions (\ref{ab}), $\varepsilon_\mathrm{I}(\mathbf{r},\omega)$ is the imaginary part of $\varepsilon(\mathbf{r},\omega)$, the relative permittivity of the desired macroscopic medium, and $\kappa_\mathrm{I}(\mathbf{r},\omega)$ is the imaginary part of $\kappa(\mathbf{r},\omega)=\mu(\mathbf{r},\omega)^{-1}$, where $\mu(\mathbf{r},\omega)$ is the relative permeability. For dissipative media, $\varepsilon_\mathrm{I}(\mathbf{r},\omega)>0$, whereas $\kappa_\mathrm{I}(\mathbf{r},\omega)<0$ because $\mu_\mathrm{I}(\mathbf{r},\omega)=-\kappa_\mathrm{I}(\mathbf{r},\omega)/|\kappa(\mathbf{r},\omega)|^2>0$. The coupling functions (\ref{ab}) are thus real and so is the action (\ref{S}); this will ensure that the Hamiltonian of the quantized theory is hermitian. The entire dielectric functions $\varepsilon(\mathbf{r},\omega)$ and $\kappa(\mathbf{r},\omega)$ of the medium, not just the imaginary parts, are specified by (\ref{ab}) because the real parts are given by the following Kramers-Kronig relation~\cite{LLcm,jac}:
\begin{equation}  \label{KK}
\varepsilon_\mathrm{R}(\mathbf{r},\omega')-1=\frac{2}{\pi}\mathrm{P}\int_0^\infty\rmd\omega\frac{\omega\varepsilon_\mathrm{I}(\mathbf{r},\omega)}{\omega^2-\omega'^2},  \qquad \mbox{and similarly for $\kappa(\mathbf{r},\omega)$.}
\end{equation}
It should also be remembered in what follows that $\varepsilon_\mathrm{I}(\mathbf{r},\omega)$ and $\kappa_\mathrm{I}(\mathbf{r},\omega)$ are odd functions of $\omega$, whereas $\varepsilon_\mathrm{R}(\mathbf{r},\omega)$ and $\kappa_\mathrm{R}(\mathbf{r},\omega)$ are even, so that  $\varepsilon^*(\mathbf{r},\omega)=\varepsilon(\mathbf{r},-\omega)$ and $\kappa ^*(\mathbf{r},\omega)= \kappa(\mathbf{r},-\omega)$, which is a consequence of the electric and magnetic susceptibilities being real~\cite{LLcm}. We have assumed the medium is isotropic, with scalar dielectric functions $\varepsilon(\mathbf{r},\omega)$ and $\kappa(\mathbf{r},\omega)$; anisotropy can be included by obvious modifications, but this is not done here as it would further tax an already burdened notation in what follows.

The coupling of electromagnetic fields to a reservoir has been considered previously in~\cite{bha06,khe06,sut07,amo08,amo09,khe10}. Here we prove that the specific action (\ref{S})--(\ref{ab}) gives the macroscopic Maxwell equations for the linear, isotropic, inhomogeneous medium with dielectric functions $\varepsilon(\mathbf{r},\omega)$ and $\kappa(\mathbf{r},\omega)$. Variation of $\phi$, $\mathbf{A}$, $\mathbf{X}_\omega$ and $\mathbf{Y}_\omega$ in the action (\ref{S}) gives, respectively,
\begin{eqnarray}
\varepsilon_0\nabla\cdot\mathbf{E}+\int_0^\infty\rmd\omega\,\nabla\cdot\left[\alpha(\mathbf{r},\omega)\mathbf{X}_\omega\right]=0,   \label{Eeq} \\[3pt]
-\kappa_0\nabla\times\mathbf{B}+\varepsilon_0\partial_t\mathbf{E}+\int_0^\infty\rmd\omega\left\{\alpha(\mathbf{r},\omega) \partial_t\mathbf{X}_\omega+\nabla\times\left[\beta(\mathbf{r},\omega)\mathbf{Y}_\omega\right]\right\}=0,  \label{Beq} \\[3pt]
-\partial_t^2\mathbf{X}_\omega-\omega^2\mathbf{X}_\omega+\alpha(\mathbf{r},\omega)\mathbf{E}=0,  \label{Xeq}  \\
-\partial_t^2\mathbf{Y}_\omega-\omega^2\mathbf{Y}_\omega+\beta(\mathbf{r},\omega)\mathbf{B}=0.  \label{Yeq}
\end{eqnarray}
In order to find the independent equations satisfied by the electromagnetic fields we must solve (\ref{Xeq}) and (\ref{Yeq}) for the reservoir fields $\mathbf{X}_\omega$ and $\mathbf{Y}_\omega$ in terms of $\mathbf{E}$ and $\mathbf{B}$, and substitute the results into (\ref{Eeq}) and (\ref{Beq}). To solve (\ref{Xeq}) we write it in the frequency domain using (\ref{Efreq}) and obtain
\begin{equation}   \label{Xeqfreq}
\omega'^2\mathbf{X}_\omega(\mathbf{r},\omega')-\omega^2\mathbf{X}_\omega(\mathbf{r},\omega') +\alpha(\mathbf{r},\omega)\mathbf{E}(\mathbf{r},\omega') =0.
\end{equation}
The general solution of (\ref{Xeqfreq}) is
\begin{equation}   \label{Xeqfreqsol}
\fl
\mathbf{X}_\omega(\mathbf{r},\omega')=\frac{\alpha(\mathbf{r},\omega)}{\omega^2-(\omega'+\rmi \omega0^+)^2}\mathbf{E}(\mathbf{r},\omega')+\delta(\omega-\omega')\mathbf{Z}_\omega(\mathbf{r})+\delta(\omega+\omega')\mathbf{Z}^*_\omega(\mathbf{r}),
\end{equation}
where $\mathbf{Z}_\omega(\mathbf{r})$ is an arbitrary function independent of $\omega'$ that gives the solution of the homogeneous equation ((\ref{Xeq}) with $\mathbf{E}=0$), and $0^+$ is an infinitesimal positive number introduced to deal with the pole at $\omega'=\omega$ that arises from dividing (\ref{Xeqfreq}) by $\omega'^2-\omega^2$. The prescription chosen in (\ref{Xeqfreqsol}) for dealing with the pole amounts to a choice of boundary condition for the reservoir field $\mathbf{X}_\omega$; in the microscopic models~\cite{hut92,sut04a,sut04b}, which also include a reservoir, similar poles are encountered in solving the coupled system of field equations and a choice is required that leads to the desired class of solutions. Note that the prescription for the pole, and the form of the solution of the homogeneous equation, must respect the frequency-domain property $\mathbf{X}^*_\omega(\mathbf{r},\omega')=\mathbf{X}_\omega(\mathbf{r},-\omega')$ that follows from the reality of the field $\mathbf{X}_\omega(\mathbf{r},t)$ in the time domain; this property holds in (\ref{Xeqfreqsol}). In transforming (\ref{Xeqfreqsol}) to the time domain the integration over frequency is determined in terms of principle values~\cite{abl} by
\begin{eqnarray}
\fl
&\frac{1}{\omega^2-(\omega'+\rmi \omega0^+)^2}=\frac{1}{2\omega}\left[\frac{1}{\omega-\omega'-\rmi \omega'0^+}+\frac{1}{\omega+\omega'-\rmi \omega'0^+}\right]  \nonumber \\[3pt]
\fl
&\qquad \qquad =\frac{1}{2\omega}\left[\mathrm{P}\frac{1}{\omega-\omega'}+\mbox{sgn}(\omega')\rmi\pi\delta(\omega-\omega')+\mathrm{P}\frac{1}{\omega+\omega'}+\mbox{sgn}(\omega')\rmi\pi\delta(\omega+\omega')\right] \nonumber \\[3pt]
 \fl
& \qquad \qquad =\mathrm{P}\frac{1}{\omega^2-\omega'^2}+\frac{\rmi\pi}{2\omega}[\delta(\omega-\omega')-\delta(\omega+\omega')] \qquad (\omega>0). \label{Ps}
\end{eqnarray}
The restriction to $\omega>0$ in the last line is appropriate because $\omega$ is positive in (\ref{Xeqfreqsol}) (see the action (\ref{S})--(\ref{Sint})). Note again that after insertion of (\ref{Ps}) in (\ref{Xeqfreqsol}) the property $\mathbf{X}^*_\omega(\mathbf{r},\omega')=\mathbf{X}_\omega(\mathbf{r},-\omega')$ is seen to hold. Using (\ref{Efreq}) in the case of $\mathbf{X}_\omega$, we obtain from (\ref{Xeqfreqsol}) and (\ref{Ps})
\begin{eqnarray}
\fl
\mathbf{X}_\omega(\mathbf{r},t)=&\frac{1}{2\pi}\mathrm{P}\int_0^\infty \rmd\omega'\left[\frac{\alpha(\mathbf{r},\omega)}{\omega^2-\omega'^2}\mathbf{E}(\mathbf{r},\omega')\exp(-\rmi\omega' t)+\mbox{c.c.}\right]  \nonumber \\[3pt]
\fl
&+\left[\frac{\rmi\alpha(\mathbf{r},\omega)}{4\omega}\mathbf{E}(\mathbf{r},\omega)\exp(-\rmi\omega t)+\mbox{c.c.}\right]+\left[\frac{1}{2\pi}\mathbf{Z}_\omega(\mathbf{r})\exp(-\rmi\omega t)+\mbox{c.c.}\right].  \label{Xsol}
\end{eqnarray}
To evaluate the $\mathbf{X}_\omega$-dependent terms in the field equations (\ref{Eeq}) and (\ref{Beq}) we require the following integral, which we evaluate using (\ref{ab}) and the Kramers-Kronig relation (\ref{KK}):
\begin{eqnarray}
\fl
\int_0^\infty\rmd\omega\,\alpha(\mathbf{r},\omega)\mathbf{X}_\omega(\mathbf{r},t)&=&\frac{\varepsilon_0}{\pi^2}\mathrm{P}\int_0^\infty \rmd\omega\int_0^\infty \rmd\omega'\left[\frac{\omega\varepsilon_\mathrm{I}(\mathbf{r},\omega)}{\omega^2-\omega'^2}\mathbf{E}(\mathbf{r},\omega')\exp(-\rmi\omega' t)+\mbox{c.c.}\right]    \nonumber \\[3pt]
\fl
&&+\frac{\varepsilon_0}{2\pi}\int_0^\infty \rmd\omega\left[\rmi\varepsilon_\mathrm{I}(\mathbf{r},\omega)\mathbf{E}(\mathbf{r},\omega)\exp(-\rmi\omega t)+\mbox{c.c.}\right]   \nonumber \\[3pt]
\fl
&&+\frac{1}{2\pi}\int_0^\infty \rmd\omega\left[\alpha(\mathbf{r},\omega)\mathbf{Z}_\omega(\mathbf{r})\exp(-\rmi\omega t)+\mbox{c.c.}\right]  \nonumber \\[3pt]
\fl
&=&-\varepsilon_0\mathbf{E}(\mathbf{r},t)+\frac{\varepsilon_0}{2\pi}\int_0^\infty \rmd\omega\left[\varepsilon(\mathbf{r},\omega)\mathbf{E}(\mathbf{r},\omega)\exp(-\rmi\omega t)+\mbox{c.c.}\right]   \nonumber \\[3pt]
\fl
&&+\frac{1}{2\pi}\int_0^\infty \rmd\omega\left[\alpha(\mathbf{r},\omega)\mathbf{Z}_\omega(\mathbf{r})\exp(-\rmi\omega t)+\mbox{c.c.}\right] . \label{alX}
\end{eqnarray}
The second term on the right-hand side of the final equality in (\ref{alX}) is the electric displacement $\mathbf{D}(\mathbf{r},t)$:
\begin{equation}  \label{D}
\mathbf{D}(\mathbf{r},t)=\frac{\varepsilon_0}{2\pi}\int_0^\infty \rmd\omega\left[\varepsilon(\mathbf{r},\omega)\mathbf{E}(\mathbf{r},\omega)\exp(-\rmi\omega t)+\mbox{c.c.}\right].
\end{equation}
Equation (\ref{Yeq}) for $\mathbf{Y}_\omega$ can be solved in an identical manner to that used to obtain (\ref{Xsol}); denoting the arbitrary function, corresponding to $\mathbf{Z}_\omega(\mathbf{r})$ in (\ref{Xsol}), by  $\mathbf{W}_\omega(\mathbf{r})$, the solution for  $\mathbf{Y}_\omega (\mathbf{r},t)$ is
\begin{eqnarray}
\fl
\mathbf{Y}_\omega(\mathbf{r},t)=&\frac{1}{2\pi}\mathrm{P}\int_0^\infty \rmd\omega'\left[\frac{\beta(\mathbf{r},\omega)}{\omega^2-\omega'^2}\mathbf{B}(\mathbf{r},\omega')\exp(-\rmi\omega' t)+\mbox{c.c.}\right]  \nonumber \\[3pt]
\fl
&+\left[\frac{\rmi\beta(\mathbf{r},\omega)}{4\omega}\mathbf{B}(\mathbf{r},\omega)\exp(-\rmi\omega t)+\mbox{c.c.}\right]+\left[\frac{1}{2\pi}\mathbf{W}_\omega(\mathbf{r})\exp(-\rmi\omega t)+\mbox{c.c.}\right].  \label{Ysol}
\end{eqnarray}
To evaluate the $\mathbf{Y}_\omega$-dependent terms in the field equations (\ref{Eeq}) and (\ref{Beq}) we require the following integral, which is analogous to (\ref{alX}) and is evaluated in an identical manner:
\begin{eqnarray}
\fl
\int_0^\infty\rmd\omega\,\beta(\mathbf{r},\omega)\mathbf{Y}_\omega(\mathbf{r},t)&=&\kappa_0\mathbf{B}(\mathbf{r},t)-\frac{\kappa_0}{2\pi}\int_0^\infty \rmd\omega\left[\kappa(\mathbf{r},\omega)\mathbf{B}(\mathbf{r},\omega)\exp(-\rmi\omega t)+\mbox{c.c.}\right]   \nonumber \\[3pt]
\fl
&&+\frac{1}{2\pi}\int_0^\infty \rmd\omega\left[\beta(\mathbf{r},\omega)\mathbf{W}_\omega(\mathbf{r})\exp(-\rmi\omega t)+\mbox{c.c.}\right] . \label{beY}
\end{eqnarray}
The second term on the right-hand side of (\ref{beY}) is the negative of the magnetic $\mathbf{H}(\mathbf{r},t)$ field:
\begin{equation} \label{Hfield}
\mathbf{H}(\mathbf{r},t)=\frac{\kappa_0}{2\pi}\int_0^\infty \rmd\omega\left[\kappa(\mathbf{r},\omega)\mathbf{B}(\mathbf{r},\omega)\exp(-\rmi\omega t)+\mbox{c.c.}\right].
\end{equation}
Substitution of (\ref{alX}),  (\ref{D}), (\ref{beY}) and (\ref{Hfield}) into the electromagnetic field equations (\ref{Eeq}) and (\ref{Beq}) yields the macroscopic Maxwell equations
\begin{eqnarray}
\nabla\cdot\mathbf{D}=\sigma, \label{gauss} \\
\nabla\times\mathbf{H}-\partial_t\mathbf{D}=\mathbf{j},  \label{amp}
\end{eqnarray}
where the free charge density $\sigma$ and free current density $\mathbf{j}$ are given by
\begin{eqnarray}
\sigma(\mathbf{r},t)=&-\frac{1}{2\pi}\nabla\cdot\int_0^\infty \rmd\omega\left[\alpha(\mathbf{r},\omega)\mathbf{Z}_\omega(\mathbf{r})\exp(-\rmi\omega t)+\mbox{c.c.}\right],  \label{sigma} \\[3pt]
\mathbf{j}(\mathbf{r},t)=&\frac{1}{2\pi}\partial_t\int_0^\infty \rmd\omega\left[\alpha(\mathbf{r},\omega)\mathbf{Z}_\omega(\mathbf{r})\exp(-\rmi\omega t)+\mbox{c.c.}\right]  \nonumber \\[3pt]
&+\frac{1}{2\pi}\nabla\times\int_0^\infty \rmd\omega\left[\beta(\mathbf{r},\omega)\mathbf{W}_\omega(\mathbf{r})\exp(-\rmi\omega t)+\mbox{c.c.}\right].
\end{eqnarray}
The charge and current densities automatically obey the conservation law
\begin{equation}  \label{conservation}
\partial_t\sigma+ \nabla\cdot \mathbf{j}=0,
\end{equation}
and the remaining two Maxwell equations are also identities because of (\ref{pots}). We have thus proven that that (\ref{S})--(\ref{Sint}) is the action for macroscopic electromagnetism.

The electric and magnetic fields are found in the usual manner. Because of (\ref{pots}), a time derivative of (\ref{amp}) yields the following frequency-domain equation for $\mathbf{E} (\mathbf{r},\omega)$:
\begin{eqnarray} 
\nabla\times\left[\kappa(\mathbf{r},\omega)\nabla\times\mathbf{E}(\mathbf{r},\omega)\right]-\frac{\omega^2}{c^2}\varepsilon(\mathbf{r},\omega)\mathbf{E}(\mathbf{r},\omega)=\rmi\mu_0\omega\mathbf{j}(\mathbf{r},\omega),  \label{Efreqeq}\\[3pt]
\mathbf{j}(\mathbf{r},\omega)=-\rmi\omega\alpha(\mathbf{r},\omega)\mathbf{Z}_\omega(\mathbf{r})+\nabla\times[\beta(\mathbf{r},\omega)\mathbf{W}_\omega(\mathbf{r})].  \label{jfreq}
\end{eqnarray}
The electric field can then be written as
\begin{equation} \label{EG}
\mathbf{E}(\mathbf{r},t)=\frac{\mu_0}{2\pi}\int_0^\infty\rmd\omega\int\rmd^3\mathbf{r'}\left[\rmi\omega\mathbf{G}(\mathbf{r},\mathbf{r'}, \omega)\cdot\mathbf{j}(\mathbf{r'},\omega)\exp(-\rmi\omega t)+\mbox{c.c}\right],
\end{equation}
where the Green bi-tensor $\mathbf{G}(\mathbf{r},\mathbf{r'}, \omega)$ is the solution of
\begin{equation} \label{green}
\nabla\times\left[\kappa(\mathbf{r},\omega)\nabla\times\mathbf{G}(\mathbf{r},\mathbf{r'}, \omega)\right]-\frac{\omega^2}{c^2}\varepsilon(\mathbf{r},\omega)\mathbf{G}(\mathbf{r},\mathbf{r'}, \omega)=\mathds{1}\delta(\mathbf{r}-\mathbf{r'})
\end{equation}
with retarded boundary conditions. Due to (\ref{pots}) the magnetic field is given in terms of $\mathbf{E}(\mathbf{r},\omega)$ by $\mathbf{B}(\mathbf{r},\omega)=-\rmi\nabla\times\mathbf{E}(\mathbf{r},\omega)/\omega$.

\section{Quantization} \label{sec:quantization}
The canonical formalism~\cite{wei} is readily applied to the action (\ref{S})--(\ref{Sint}). A degree of choice is present in writing the Lagrangian density $\mathcal{L}$, associated with the action (\ref{S}) by
\begin{equation}  \label{L}
S[\phi,\mathbf{A},\mathbf{X}_\omega,\mathbf{Y}_\omega]=\int\rmd^4x\,\mathcal{L},
\end{equation}
since integrations by parts in (\ref{L}) change the form of $\mathcal{L}$ without affecting the dynamics. We take $\mathcal{L}$ to be given by the integrands (\ref{S})--(\ref{Sint}), without any integrations by parts. The canonical momenta are then
\begin{eqnarray}
\Pi_\phi=\frac{\partial\mathcal{L}}{\partial(\partial_t\phi)}=0, \qquad \mathbf{\Pi}_\mathrm{A}=\frac{\partial\mathcal{L}}{\partial(\partial_t\mathbf{A})}=-\varepsilon_0\mathbf{E}-\int_0^\infty \rmd\omega\,\alpha(\mathbf{r},\omega)\mathbf{X}_\omega, \label{mom1} \\[3pt]
\mathbf{\Pi}_{\mathrm{X}_\omega}=\frac{\partial\mathcal{L}}{\partial(\partial_t\mathbf{\mathbf{X}_\omega})}=\partial_t\mathbf{X}_\omega, \qquad  \mathbf{\Pi}_{\mathrm{Y}_\omega}=\frac{\partial\mathcal{L}}{\partial(\partial_t\mathbf{\mathbf{Y}_\omega})}=\partial_t\mathbf{Y}_\omega.  \label{mom2}
\end{eqnarray}
The familiar complications of canonical QED~\cite{wei} are present here: the canonical momentum associated with the scalar potential $\phi$ vanishes and the gauge freedom also reduces the number of independent electromagnetic degrees of freedom. These difficulties are dealt with~\cite{wei} by eliminating $\phi$ as a dynamical variable, replacing it by its solution obtained from the field equations, and by making a choice of gauge. Even after $\phi$ is removed from the list of canonical variables, the vanishing of $\Pi_\phi$ is responsible~\cite{wei} for a constraint on $\mathbf{\Pi}_ \mathrm{A}$: from the field equation (\ref{Eeq}) and the expression for $\mathbf{\Pi}_ \mathrm{A}$ in (\ref{mom1}) we obtain
\begin{equation} \label{con1}
\nabla\cdot\mathbf{\Pi}_ \mathrm{A}=0.
\end{equation}
The general method for quantizing field theories with constraints, such as (\ref{con1}) and a choice of gauge, is that of Dirac brackets~\cite{wei}. For unconstrained theories the classical Poisson brackets of the canonical variables are to be replaced by $-\rmi/\hbar$ times quantum commutators, but when there are constraints on the canonical variables it is a generalization of the canonical Poisson brackets, known as Dirac brackets, that become commutators  in the quantized theory. The Dirac brackets are computed from the Poisson brackets of the quantities that vanish due to the constraints~\cite{wei}; this requires that the constraints be written in terms of the canonical variables. Here (\ref{con1}) is one constraint, and a choice of gauge will give another. For free-space QED the Coulomb gauge
\begin{equation}  \label{gauge}
\nabla\cdot \mathbf{A}=0
\end{equation}
is a convenient choice, but the macroscopic Maxwell equation (\ref{gauss}) suggests a generalization that has the frequency-domain form 
\begin{equation} \label{gaugenot}
\nabla\cdot[\varepsilon(\mathbf{r},\omega)\mathbf{A} (\mathbf{r},\omega)]=0.
\end{equation}
In homogeneous media (\ref{gaugenot}) reduces to the Coulomb gauge (\ref{gauge}), but the gauge choice (\ref{gaugenot}) is not appropriate for inhomogeneous media because in the time domain it cannot be written in terms of the dynamical fields and their canonical momenta. As a result, Dirac brackets involving the constraint (\ref{gaugenot}) cannot be calculated and there is no obvious way to proceed with the quantization. We therefore choose the Coulomb gauge (\ref{gauge}), even in the general case of inhomogeneous media.

After the elimination of $\phi$ as a dynamical field and imposition of the Coulomb gauge, the electromagnetic canonical variables are $\mathbf{A}$ and $\mathbf{\Pi}_ \mathrm{A}$, subject to the constraints (\ref{con1}) and (\ref{gauge}). These are the same constraints as in free-space QED in the Coulomb gauge~\cite{wei}; an evaluation of the Dirac brackets arising from these constraints shows~\cite{wei} that the quantum canonical operators $\mathbf{\hat{A}}$ and $\mathbf{\hat{\Pi}}_ \mathrm{A}$ should be defined by the equal-time commutation relations
\begin{equation} \label{comem}
\eqalign{
[\hat{A}_i(\mathbf{r},t),\hat{\Pi}_{\mathrm{A}j} (\mathbf{r'},t)]=\rmi\hbar\delta_{\mathrm{T}ij}(\mathbf{r}-\mathbf{r'}), \cr
[\hat{A}_i(\mathbf{r},t),\hat{A}_j(\mathbf{r'},t)]=0, \qquad [\hat{\Pi}_{\mathrm{A}i} (\mathbf{r},t),\hat{\Pi}_{\mathrm{A}j} (\mathbf{r'},t)]=0.
}
\end{equation}
where $\delta_{\mathrm{T}ij}(\mathbf{r}-\mathbf{r'})$ is the transverse delta function~\cite{wei,lou}
\begin{equation}
\delta_{\mathrm{T}ij}(\mathbf{r}-\mathbf{r'})=\delta_{ij}\delta(\mathbf{r}-\mathbf{r'})+\partial_i \partial_j\left(\frac{1}{4\pi|\mathbf{r}-\mathbf{r'}|}\right).
\end{equation}
The canonical operators of the reservoir are subject to no constraints and therefore satisfy the commutation relations
\begin{eqnarray}
[\hat{X}_{\omega i}(\mathbf{r},t),\hat{\Pi}_{\mathrm{X}_{\omega'} j} (\mathbf{r'},t)]=\rmi\hbar\delta_{ij}\delta(\omega-\omega')\delta(\mathbf{r}-\mathbf{r'}), \\ 
{[}\hat{X}_{\omega i}(\mathbf{r},t),\hat{X}_{\omega' j}(\mathbf{r'},t)]=0, \qquad [\hat{\Pi}_{\mathrm{X}_\omega i} (\mathbf{r},t),\hat{\Pi}_{\mathrm{X}_{\omega'} j} (\mathbf{r'},t)]=0, \\
{[}\hat{Y}_{\omega i}(\mathbf{r},t),\hat{\Pi}_{\mathrm{Y}_{\omega'} j} (\mathbf{r'},t)]=\rmi\hbar\delta_{ij}\delta(\omega-\omega')\delta(\mathbf{r}-\mathbf{r'}), \\ 
{[}\hat{Y}_{\omega i}(\mathbf{r},t),\hat{Y}_{\omega' j}(\mathbf{r'},t)]=0, \qquad [\hat{\Pi}_{\mathrm{Y}_\omega i} (\mathbf{r},t),\hat{\Pi}_{\mathrm{Y}_{\omega'} j} (\mathbf{r'},t)]=0.   \label{comY}
\end{eqnarray}
Both $\mathbf{\hat{A}}$ and $\mathbf{\hat{\Pi}}_ \mathrm{A}$ are transverse due to the constraints (\ref{gauge}) and (\ref{con1}) and this is consistent with the commutation relations (\ref{comem}). From (\ref{pots}), (\ref{mom1}) and (\ref{con1}) we see that the transverse part of the electric field $\mathbf{\hat{E}}$ is equal to $-\partial_t\mathbf{\hat{A}}$, while the longitudinal part of $\mathbf{\hat{E}}$ determines the scalar potential $\hat{\phi}$ and is given by the longitudinal part of $\int_0^\infty \rmd\omega\,\alpha(\mathbf{r},\omega)\mathbf{\hat{X}}_\omega$:
\begin{eqnarray}
\mathbf{\hat{E}}=\mathbf{\hat{E}}_\mathrm{T}+\mathbf{\hat{E}}_\mathrm{L}, \\
\mathbf{\hat{E}}_\mathrm{T}=-\partial_t\mathbf{\hat{A}}, \qquad
\mathbf{\hat{E}}_\mathrm{L}=-\nabla\hat{\phi}=-\varepsilon_0^{-1}\left[\int_0^\infty \rmd\omega\,\alpha(\mathbf{r},\omega)\mathbf{\hat{X}}_\omega\right]_\mathrm{L}. \label{ETL}
\end{eqnarray}

To write the Hamiltonian we must express the time derivatives of the dynamical fields $\mathbf{\hat{A}}$, $\mathbf{\hat{X}}_\omega$ and $\mathbf{\hat{Y}}_\omega$ in terms of their canonical momenta; this is done for the reservoir fields by (\ref{mom2}), while for the vector potential we obtain from (\ref{mom1}) and (\ref{pots})
\begin{equation}  \label{dtA}
\partial_t\mathbf{\hat{A}}=\varepsilon_0^{-1}\left[\mathbf{\hat{\Pi}}_ \mathrm{A}-\varepsilon_0\nabla\hat{\phi}+\int_0^\infty \rmd\omega\,\alpha(\mathbf{r},\omega)\mathbf{\hat{X}}_\omega\right].
\end{equation}
The Hamiltonian density $\hat{\mathcal{H}}$ is given by
\begin{equation} \label{Hden}
\hat{\mathcal{H}}=\mathbf{\hat{\Pi}}_ \mathrm{A}\cdot\partial_t\mathbf{\hat{A}}+\mathbf{\hat{\Pi}}_{\mathrm{X}_\omega}\cdot\partial_t\mathbf{\hat{X}}_\omega+\mathbf{\hat{\Pi}}_{\mathrm{Y}_\omega}\cdot \partial_t\mathbf{\hat{Y}}_\omega-\hat{\mathcal{L}}.
\end{equation}
To express the Hamiltonian $\hat{H}=\int\rmd^3\mathbf{r}\,\hat{\mathcal{H}}$ in terms of the canonical variables, all time derivatives in (\ref{Hden}) must be eliminated using (\ref{mom2}) and (\ref{dtA}). After substitution of (\ref{dtA}) into the first term of the Hamiltonian density (\ref{Hden}) we encounter a term $-\mathbf{\hat{\Pi}}_ \mathrm{A}\cdot\nabla\hat{\phi}$; upon taking the volume integral $\int\rmd^3\mathbf{r}\,\hat{\mathcal{H}}$ this term becomes, after an integration by parts, $\int\rmd^3\mathbf{r}\,\hat{\phi}\nabla \cdot\mathbf{\hat{\Pi}}_ \mathrm{A}$, which vanishes by (\ref{con1}). Alternatively, the term $-\int\rmd^3\mathbf{\hat{\Pi}}_ \mathrm{A}\cdot\nabla\hat{\phi}$ can be seen to vanish by noting that $\nabla\hat{\phi}$ is a longitudinal vector, whereas $\mathbf{\hat{\Pi}}_ \mathrm{A}$ is transverse by (\ref{con1}); the volume integration of the scalar product of a transverse with a longitudinal field necessarily vanishes~\cite{lou}. In the Lagrangian-density term in (\ref{Hden}) the electric field can be  expressed in term of the canonical variables by (\ref{mom1}) and the magnetic field by (\ref{pots}). The resulting Hamiltonian is
\begin{eqnarray}
\fl
\hat{H}=\int\rmd^3\mathbf{r}&\left\{\frac{1}{\varepsilon_0}\mathbf{\hat{\Pi}}_ \mathrm{A}\cdot\left[\frac{1}{2}\mathbf{\hat{\Pi}}_ \mathrm{A}+\int_0^\infty \rmd\omega\,\alpha(\mathbf{r},\omega)\mathbf{\hat{X}}_\omega\right]+\frac{\kappa_0}{2}(\nabla\times\mathbf{\hat{A}})^2\right. \nonumber \\[3pt]
\fl
&\ \ +\frac{1}{2}\int_0^\infty\rmd\omega\left[\left(\mathbf{\hat{\Pi}}_{\mathrm{X}_\omega}^2+\mathbf{\hat{\Pi}}_{\mathrm{Y}_\omega}^2\right)+\omega^2\left(\mathbf{\hat{X}}_\omega^2+\mathbf{\hat{Y}}_\omega^2\right)\right]   \nonumber  \\[3pt]
\fl
&\ \ +\frac{1}{2 \varepsilon_0}\left[\int_0^\infty \rmd\omega\,\alpha(\mathbf{r},\omega)\mathbf{\hat{X}}_\omega\right]\cdot\left[\int_0^\infty \rmd\omega'\,\alpha(\mathbf{r},\omega')\mathbf{\hat{X}}_{\omega'}\right]  \nonumber  \\[3pt]
\fl
&\ \ -\left.\int_0^\infty \rmd\omega\,\beta(\mathbf{r},\omega)\mathbf{\hat{Y}}_\omega\cdot(\nabla\times\mathbf{\hat{A}})\right\}.  \label{H}
\end{eqnarray}

The Hamiltonian (\ref{H}) gives dynamical equations for the field operators that are the quantum versions of the classical equations (\ref{Eeq})--(\ref{Yeq}). First note that (\ref{Eeq}) is now a statement of the constraint (\ref{con1}) (see (\ref{mom1})), rather than a dynamical equation. The Hamiltonian (\ref{H}) and the commutation relations (\ref{comem}) determine the time derivative of $\mathbf{\hat{A}}$ as
\begin{equation} \label{AH}
\partial_t\mathbf{\hat{A}}=-\frac{\rmi}{\hbar}[\mathbf{\hat{A}},\hat{H}]=\frac{1}{\varepsilon_0}\left[\mathbf{\hat{\Pi}}_ \mathrm{A}+\left(\int_0^\infty \rmd\omega\,\alpha(\mathbf{r},\omega)\mathbf{\hat{X}}_\omega\right)_\mathrm{T}\right].
\end{equation}
In this equation the transverse part of $\int_0^\infty \rmd\omega\,\alpha(\mathbf{r},\omega)\mathbf{\hat{X}}_\omega$ appears because its longitudinal part does not contribute the volume integral of $\mathbf{\hat{\Pi}}_ \mathrm{A}\cdot\int_0^\infty \rmd\omega\,\alpha(\mathbf{r},\omega)\mathbf{\hat{X}}_\omega$ in (\ref{H}) (recall that $\mathbf{\hat{\Pi}}_ \mathrm{A}$ is transverse). The time derivative of $\mathbf{\hat{\Pi}}_ \mathrm{A}$ is
\begin{equation}  \label{PAH}
\partial_t\mathbf{\hat{\Pi}}_ \mathrm{A}=-\frac{\rmi}{\hbar}[\mathbf{\hat{\Pi}}_ \mathrm{A},\hat{H}]=-\kappa_0\nabla\times(\nabla\times\mathbf{\hat{A}})+\nabla\times\int_0^\infty \rmd\omega\,\beta(\mathbf{r},\omega)\mathbf{\hat{Y}}_\omega.
\end{equation}
Eliminating $\mathbf{\hat{\Pi}}_ \mathrm{A}$ from (\ref{AH}) and (\ref{PAH}) gives the dynamical equation for $\mathbf{\hat{A}}$, which can be written in terms of the transverse part of $\mathbf{\hat{E}}$ by (\ref{ETL}):
\begin{equation}
\fl
-\varepsilon_0\partial_t\mathbf{\hat{E}}_\mathrm{T}-\partial_t\left(\int_0^\infty \rmd\omega\,\alpha(\mathbf{r},\omega)\mathbf{\hat{X}}_\omega\right)_\mathrm{T}=-\kappa_0\nabla\times\mathbf{\hat{B}}+\nabla\times\int_0^\infty \rmd\omega\,\beta(\mathbf{r},\omega)\mathbf{\hat{Y}}_\omega.
\end{equation}
This is simply the quantum version of the transverse part of (\ref{Beq}); the longitudinal part of (\ref{Beq}) appears here as the time derivative of the second equation in (\ref{ETL}), which arises from the constraints. The full equation (\ref{Beq}) thus holds also in the quantum theory. It is straightforward to verify, in a similar manner, that the Hamiltonian (\ref{H}) gives dynamical equations for $\mathbf{\hat{X}}_\omega$ and $\mathbf{\hat{Y}}_\omega$ that are simply the quantum versions of (\ref{Xeq}) and (\ref{Yeq}), respectively. It therefore follows from the results of the last Section that the quantum theory specified by the commutation relations (\ref{comem})--(\ref{comY}) and the Hamiltonian (\ref{H}) gives the quantum macroscopic Maxwell equations (\ref{gauss})--(\ref{amp}) for $\mathbf{\hat{D}}$ and $\mathbf{\hat{H}}$. These quantum Maxwell equations will also be explicitly derived in the following Section.

\section{Diagonalization of the Hamiltonian} \label{sec:diag}
We now proceed to diagonalize the Hamiltonian (\ref{H}), which essentially amounts to solving the Hamilton equations for the quantum canonical operators. This diagonalization is a familiar procedure from quantization of the microscopic models~\cite{hut92,sut04a,sut04b,khe08}, and we follow the method used in~\cite{sut04b}, where the model was of an inhomogeneous dielectric. The field operators in the diagonalized Hamiltonian will be bosonic creation and annihilation operators for the eigenmodes of the system. We will show that the Hamiltonian can be written
\begin{equation} \label{Hdiag}
\hat{H}=\sum_{\lambda=\mathrm{e},\mathrm{m}}\int\rmd^3 \mathbf{r}\int_0^\infty\rmd\omega\,\hbar\omega\mathbf{\hat{C}}^\dagger_\lambda(\mathbf{r},\omega)\cdot\mathbf{\hat{C}}_\lambda (\mathbf{r},\omega),
\end{equation}
where $\mathbf{\hat{C}}_\mathrm{e}(\mathbf{r},\omega)$ and $\mathbf{\hat{C}}_\mathrm{m}(\mathbf{r},\omega)$ are annihilation operators associated with the electric and magnetic responses of the medium, respectively. The diagonalizing operators obey the usual commutation relations of bosonic annihilation and creation operators:
\begin{equation} \label{CC}
\fl
\left[\hat{C}_{\lambda i}(\mathbf{r},\omega),\hat{C}^\dagger_{\lambda' j}(\mathbf{r'},\omega')\right]=\delta_{ij}\delta_{\lambda\lambda'} \delta(\omega-\omega') \delta(\mathbf{r}-\mathbf{r'}), \qquad \left[\hat{C}_{\lambda i}(\mathbf{r},\omega),\hat{C}_{\lambda' j}(\mathbf{r'},\omega')\right]=0.
\end{equation}
From (\ref{Hdiag}) and (\ref{CC}) we obtain the commutators
\begin{equation} \label{CH}
\left[\hat{C}_{\lambda i}(\mathbf{r},\omega),\hat{H}\right]=\hbar\omega\mathbf{\hat{C}}_\lambda(\mathbf{r},\omega),
\end{equation}
which imply that the  time-dependent operators are given by
\begin{equation} \label{Ct}
\mathbf{\hat{C}}_\lambda(\mathbf{r},t,\omega)=\exp(-\rmi \omega t)\mathbf{\hat{C}}_\lambda(\mathbf{r},\omega).
\end{equation}
The canonical operators in the Hamiltonian (\ref{H}) must be expressible as a linear combination of the diagonalizing operators; for example, it must be possible to write the vector potential $\mathbf{\hat{A}}$ and it canonical momentum $\mathbf{\hat{\Pi}}_\mathrm{A}$ as
\begin{eqnarray} 
\mathbf{\hat{A}}(\mathbf{r},t)=\sum_{\lambda}\int\rmd^3 \mathbf{r'}\int_0^\infty\rmd\omega\, [\mathbf{f}^\lambda_\mathrm{A}(\mathbf{r},\mathbf{r'},\omega)\cdot\mathbf{\hat{C}}_\lambda(\mathbf{r'},t,\omega)+\mbox{h.c.}],   \label{AC} \\[3pt]
\mathbf{\hat{\Pi}}_\mathrm{A}(\mathbf{r},t)=\sum_{\lambda}\int\rmd^3 \mathbf{r'}\int_0^\infty\rmd\omega\, [\mathbf{f}^\lambda_{\mathrm{\Pi_A}}(\mathbf{r},\mathbf{r'},\omega)\cdot\mathbf{\hat{C}}_\lambda(\mathbf{r'},t,\omega)+\mbox{h.c.}],  \label{PAC} 
\end{eqnarray}
where $\mathbf{f}^\lambda_\mathrm{A}(\mathbf{r},\mathbf{r'},\omega)$ and $\mathbf{f}^\lambda_{\mathrm{\Pi_A}} (\mathbf{r},\mathbf{r'},\omega)$ are c-number bi-tensors. Relations similar to (\ref{AC}) and (\ref{PAC}) must also hold for the remaining canonical operators in (\ref{H}); thus for $\mathbf{\hat{X}}_\omega$ and $\mathbf{\hat{\Pi}}_\mathrm{X_ \omega}$ we require
\begin{eqnarray} 
\mathbf{\hat{X}}_\omega(\mathbf{r},t)=\sum_{\lambda}\int\rmd^3 \mathbf{r'}\int_0^\infty\rmd\omega'\, [\mathbf{f}^\lambda_{\mathrm{X}}(\mathbf{r},\mathbf{r'},\omega, \omega')\cdot\mathbf{\hat{C}}_\lambda(\mathbf{r'},t,\omega')+\mbox{h.c.}],   \label{XC} \\[3pt]
\mathbf{\hat{\Pi}}_\mathrm{X_ \omega}(\mathbf{r},t)=\sum_{\lambda}\int\rmd^3 \mathbf{r'}\int_0^\infty\rmd\omega'\, [\mathbf{f}^\lambda_{\mathrm{\Pi_X}}(\mathbf{r},\mathbf{r'},\omega,\omega')\cdot\mathbf{\hat{C}}_\lambda(\mathbf{r'},t,\omega')+\mbox{h.c.}],  \label{PXC} 
\end{eqnarray}
with analogous formulae for $\mathbf{\hat{Y}}_\omega$ and $\mathbf{\hat{\Pi}}_\mathrm{Y_ \omega}$ in terms of bi-tensorial coefficients $\mathbf{f}^\lambda_{\mathrm{Y}}$ and $\mathbf{f}^\lambda_{\mathrm{\Pi_Y}}$. The bi-tensors  
$\mathbf{f}^\lambda_\mathrm{A}$ and $\mathbf{f}^\lambda_{\mathrm{\Pi_A}}$ are transverse because of (\ref{gauge}) and (\ref{con1}). From (\ref{mom1}) we see that (\ref{AC}) and (\ref{XC}) imply an expansion of  the electric field $\mathbf{\hat{E}}$ similar to (\ref{AC}), in terms of a bi-tensorial coefficient $\mathbf{f}^\lambda_{\mathrm{E}}$ related to $\mathbf{f}^\lambda_{\mathrm{A}}$ and $\mathbf{f}^\lambda_{\mathrm{X}}$ by
\begin{equation} \label{fE}
\mathbf{f}^\lambda_\mathrm{E}(\mathbf{r},\mathbf{r'},\omega)=-\frac{1}{\varepsilon_0}\mathbf{f}^\lambda_{\mathrm{\Pi_A}}(\mathbf{r},\mathbf{r'},\omega)-\frac{1}{\varepsilon_0}\int_0^\infty \rmd\omega'\,\alpha(\mathbf{r},\omega')\mathbf{f}^\lambda_{\mathrm{X}}(\mathbf{r},\mathbf{r'},\omega', \omega).
\end{equation}
To establish the diagonalization we must show that there exist bi-tensorial coefficients in (\ref{AC})--(\ref{PXC}) (and in the corresponding equations for $\mathbf{\hat{Y}}_\omega$ and $\mathbf{\hat{\Pi}}_\mathrm{Y_ \omega}$) such that the Hamiltonian (\ref{H}) takes the form (\ref{Hdiag}).

From (\ref{CC}) and (\ref{AC})--(\ref{PXC}) it follows that the desired bi-tensorial coefficients must be given by commutators of the canonical operators with the diagonalizing operators; we give two examples:
\begin{equation} \label{fAC}
\fl
\mathbf{f}^\lambda_\mathrm{A}(\mathbf{r},\mathbf{r'},\omega)=[\mathbf{\hat{A}}(\mathbf{r},t),\mathbf{\hat{C}}^\dagger_\lambda(\mathbf{r'},t,\omega)], \qquad  \mathbf{f}^\lambda_\mathrm{X}(\mathbf{r},\mathbf{r'},\omega, \omega')=[\mathbf{\hat{X}}_\omega(\mathbf{r},t),\mathbf{\hat{C}}^\dagger_\lambda(\mathbf{r'},t,\omega')].
\end{equation}
The diagonalizing operators must in turn be expressible as linear combinations of the canonical operators (the inverse of the relations (\ref{AC})--(\ref{PXC})) and the collection of commutators of type (\ref{fAC}), together with (\ref{comem})--(\ref{comY}), imply that $\mathbf{\hat{C}}_\lambda(\mathbf{r},t,\omega)$ has the expansion
\begin{eqnarray}
\fl
\mathbf{\hat{C}}_\lambda(\mathbf{r},t,\omega)=&-\frac{\rmi}{\hbar}\int\rmd^3 \mathbf{r'}\left\{\mathbf{\hat{A}}(\mathbf{r'},t)\cdot\mathbf{f}^{\lambda*}_{\mathrm{\Pi_A}}(\mathbf{r'},\mathbf{r},\omega)-\mathbf{\hat{\Pi}}_ \mathrm{A}(\mathbf{r'},t)\cdot\mathbf{f}^{\lambda*}_{\mathrm{A}}(\mathbf{r'},\mathbf{r},\omega) \right. \nonumber  \\[3pt]
\fl
&+\int_0^\infty \rmd\omega'\left[\mathbf{\hat{X}}_{\omega'}(\mathbf{r'},t)\cdot\mathbf{f}^{\lambda*}_{\mathrm{\Pi_X}}(\mathbf{r'},\mathbf{r},\omega',\omega)- \mathbf{\hat{\Pi}}_{\mathrm{X_{\omega'}}}(\mathbf{r'},t)\cdot\mathbf{f}^{\lambda*}_{\mathrm{X}}(\mathbf{r'},\mathbf{r},\omega',\omega)\right.  \nonumber  \\[3pt]
\fl
&\left.\left.+\mathbf{\hat{Y}}_{\omega'}(\mathbf{r'},t)\cdot\mathbf{f}^{\lambda*}_{\mathrm{\Pi_Y}}(\mathbf{r'},\mathbf{r},\omega',\omega)- \mathbf{\hat{\Pi}}_{\mathrm{Y_{\omega'}}}(\mathbf{r'},t)\cdot\mathbf{f}^{\lambda*}_{\mathrm{Y}}(\mathbf{r'},\mathbf{r},\omega',\omega)\right]\right\},  \label{Ccan}
\end{eqnarray}
where the asterisks on the bi-tensorial coefficients denote complex conjugation. We can now substitute (\ref{Ccan}) and (\ref{H}) into (\ref{CH}) and use (\ref{comem})--(\ref{comY}) to obtain an expansion of $\hbar\omega\mathbf{\hat{C}}_\lambda(\mathbf{r},t,\omega)$ in terms of the canonical operators; when the coefficients in this expansion are compared with those in the expansion (\ref{Ccan}) we find the following conditions on the bi-tensorial coefficients:
\begin{eqnarray}
\fl
\rmi \omega\mathbf{f}^{\lambda}_{\mathrm{\Pi_A}}(\mathbf{r'},\mathbf{r},\omega)=\kappa_0\nabla'\times[\nabla'\times\mathbf{f}^{\lambda}_{\mathrm{A}}(\mathbf{r'},\mathbf{r},\omega)]-\int_0^\infty \rmd\omega'\nabla'\times[\beta(\mathbf{r'},\omega')\mathbf{f}^\lambda_{\mathrm{Y}}(\mathbf{r'},\mathbf{r},\omega', \omega)],    \label{feq1} \\[3pt]
\fl
\rmi \omega\mathbf{f}^{\lambda}_{\mathrm{A}}(\mathbf{r'},\mathbf{r},\omega)=-\frac{1}{\varepsilon_0}\mathbf{f}^{\lambda}_{\mathrm{\Pi_A}}(\mathbf{r'},\mathbf{r},\omega)-\frac{1}{\varepsilon_0}\left[\int_0^\infty \rmd\omega'\,\alpha(\mathbf{r'},\omega')\mathbf{f}^\lambda_{\mathrm{X}}(\mathbf{r'},\mathbf{r},\omega', \omega)\right]_\mathrm{T},  \label{feq2}  \\[3pt]
\fl
\rmi \omega\mathbf{f}^{\lambda}_{\mathrm{\Pi_X}}(\mathbf{r'},\mathbf{r}, \omega',\omega)=\frac{1}{\varepsilon_0}\alpha(\mathbf{r'},\omega')\mathbf{f}^{\lambda}_{\mathrm{\Pi_A}}(\mathbf{r'},\mathbf{r},\omega)+ \omega'^2\mathbf{f}^\lambda_{\mathrm{X}}(\mathbf{r'},\mathbf{r},\omega', \omega)  \nonumber   \\[3pt]
+\frac{1}{\varepsilon_0}\alpha(\mathbf{r'},\omega')\int_0^\infty \rmd\omega''\,\alpha(\mathbf{r'},\omega'')\mathbf{f}^\lambda_{\mathrm{X}}(\mathbf{r'},\mathbf{r},\omega'', \omega),   \label{feq3}  \\[3pt]
\fl
\rmi \omega\mathbf{f}^\lambda_{\mathrm{X}}(\mathbf{r'},\mathbf{r},\omega', \omega)=-\mathbf{f}^\lambda_{\mathrm{\Pi_X}}(\mathbf{r'},\mathbf{r},\omega', \omega),    \label{feq4}  \\
\fl
\rmi \omega\mathbf{f}^\lambda_{\mathrm{\Pi_Y}}(\mathbf{r'},\mathbf{r},\omega', \omega)=-\beta(\mathbf{r'},\omega')\nabla'\times\mathbf{f}^{\lambda}_{\mathrm{A}}(\mathbf{r'},\mathbf{r},\omega)+\omega'^2\mathbf{f}^\lambda_{\mathrm{Y}}(\mathbf{r'},\mathbf{r},\omega', \omega),   \label{feq5}  \\
\fl
\rmi \omega\mathbf{f}^\lambda_{\mathrm{Y}}(\mathbf{r'},\mathbf{r},\omega', \omega)=-\mathbf{f}^\lambda_{\mathrm{\Pi_Y}}(\mathbf{r'},\mathbf{r},\omega', \omega).   \label{feq6}
\end{eqnarray}
The last term in (\ref{feq2}) is the transverse part in respect of the first index on this bi-tensor; the transverse apart appears because this quantity was the coefficient of a transverse operator in a volume integral. Note that (\ref{feq1}) and (\ref{feq2}) are consistent with the transverseness of $\mathbf{f}^\lambda_\mathrm{A}$ and $\mathbf{f}^\lambda_{\mathrm{\Pi_A}}$, mentioned after (\ref{PXC}). We must now solve (\ref{feq1})--(\ref{feq6}) to find the required bi-tensorial coefficients.

Equations (\ref{feq3}) and (\ref{feq4}) imply
\begin{eqnarray}
\fl
\omega^2\mathbf{f}^\lambda_{\mathrm{X}}(\mathbf{r'},\mathbf{r},\omega', \omega)&=&\omega'^2\mathbf{f}^\lambda_{\mathrm{X}}(\mathbf{r'},\mathbf{r},\omega', \omega)  \nonumber \\[3pt]
\fl
&&+\frac{1}{\varepsilon_0}\alpha(\mathbf{r'},\omega')\left[\mathbf{f}^{\lambda}_{\mathrm{\Pi_A}}(\mathbf{r'},\mathbf{r},\omega)+\int_0^\infty \rmd\omega''\,\alpha(\mathbf{r'},\omega'')\mathbf{f}^\lambda_{\mathrm{X}}(\mathbf{r'},\mathbf{r},\omega'', \omega)\right]   \nonumber \\[3pt]
\fl
&=&\omega'^2\mathbf{f}^\lambda_{\mathrm{X}}(\mathbf{r'},\mathbf{r},\omega', \omega)-\alpha(\mathbf{r'},\omega')\mathbf{f}^\lambda_{\mathrm{E}}(\mathbf{r'},\mathbf{r},\omega),  \label{fXeq}
\end{eqnarray}
where (\ref{fE}) has been used in the second equality. Equation (\ref{fXeq}) is essentially the same as (\ref{Xeqfreq}), and we write its general solution with the same pole prescription as in (\ref{Xeqfreqsol}):
\begin{equation}   \label{fXsol}
\fl
\mathbf{f}^\lambda_{\mathrm{X}}(\mathbf{r'},\mathbf{r},\omega', \omega)=\frac{\alpha(\mathbf{r'},\omega')}{\omega'^2-(\omega+\rmi \omega'0^+)^2}\mathbf{f}^\lambda_{\mathrm{E}}(\mathbf{r'},\mathbf{r},\omega)+\delta(\omega-\omega')\mathbf{h}^\lambda_{\mathrm{X}}(\mathbf{r'},\mathbf{r},\omega).
\end{equation}
The last term in (\ref{fXsol}) is the solution of the homogeneous equation ((\ref{fXeq}) with $\mathbf{f}^\lambda_{\mathrm{E}}=0$) featuring an arbitrary bi-tensor $\mathbf{h}^\lambda_{\mathrm{X}}$. We do not need to include a term proportional to $\delta(\omega+\omega')$ in (\ref{fXsol}), as is done in  (\ref{Xeqfreqsol}), because we need only consider positive frequency arguments in the bi-tensorial coefficients. From (\ref{feq5}) and (\ref{feq6}) we find a solution for $\mathbf{f}^\lambda_{\mathrm{Y}}$ similar to (\ref{fXsol}):
\begin{equation}   \label{fYsol}
\fl
\mathbf{f}^\lambda_{\mathrm{Y}}(\mathbf{r'},\mathbf{r},\omega', \omega)=\frac{\beta(\mathbf{r'},\omega')}{\omega'^2-(\omega+\rmi \omega'0^+)^2}\nabla'\times\mathbf{f}^\lambda_{\mathrm{A}}(\mathbf{r'},\mathbf{r},\omega)+\delta(\omega-\omega')\mathbf{h}^\lambda_{\mathrm{Y}}(\mathbf{r'},\mathbf{r},\omega).
\end{equation}

Elimination of $\mathbf{f}^\lambda_{\mathrm{\Pi_A}}$ from (\ref{feq1}) and (\ref{feq2}) gives
\begin{eqnarray}
\fl
\varepsilon_0\omega^2\mathbf{f}^\lambda_{\mathrm{A}}(\mathbf{r'},\mathbf{r},\omega)=&\rmi\omega\left[\int_0^\infty \rmd\omega'\,\alpha(\mathbf{r'},\omega')\mathbf{f}^\lambda_{\mathrm{X}}(\mathbf{r'},\mathbf{r},\omega', \omega)\right]_\mathrm{T}+\kappa_0\nabla'\times[\nabla'\times\mathbf{f}^{\lambda}_{\mathrm{A}}(\mathbf{r'},\mathbf{r},\omega)]  \nonumber \\[3pt]
\fl
&-\int_0^\infty \rmd\omega'\nabla'\times[\beta(\mathbf{r'},\omega')\mathbf{f}^\lambda_{\mathrm{Y}}(\mathbf{r'},\mathbf{r},\omega', \omega)]. \label{fAeq}
\end{eqnarray}
The integrations in (\ref{fAeq}) can be performed using (\ref{fXsol}) and (\ref{fYsol}), in the same way as the integrals (\ref{alX}) and (\ref{beY}) were evaluated in Section~\ref{sec:action}. We use the principle-value relation (\ref{Ps}) in (\ref{fXsol}) and (\ref{fYsol}), where here the term proportional to $\delta(\omega+\omega')$ does not contribute as we take only positive frequency arguments in the bi-tensorial coefficients. The definitions (\ref{ab}) and the Kramers-Kronig relation (\ref{KK}) are also required, and we obtain
\begin{eqnarray}
\fl
\int_0^\infty \rmd\omega'\,\alpha(\mathbf{r'},\omega')\mathbf{f}^\lambda_{\mathrm{X}}(\mathbf{r'},\mathbf{r},\omega', \omega)=\alpha(\mathbf{r'},\omega)\mathbf{h}^\lambda_{\mathrm{X}}(\mathbf{r'},\mathbf{r},\omega)+ \varepsilon_0[\varepsilon(\mathbf{r'},\omega)-1]\mathbf{f}^\lambda_{\mathrm{E}}(\mathbf{r'},\mathbf{r},\omega),  \label{alfX} \\[3pt]
\fl
\int_0^\infty\!\!\!\!\rmd\omega'\,\beta(\mathbf{r'},\omega')\mathbf{f}^\lambda_{\mathrm{Y}}(\mathbf{r'},\mathbf{r},\omega', \omega)=\beta(\mathbf{r'},\omega)\mathbf{h}^\lambda_{\mathrm{Y}}(\mathbf{r'},\mathbf{r},\omega) -\kappa_0[\kappa(\mathbf{r'},\omega)-1]\nabla'\times\mathbf{f}^\lambda_{\mathrm{A}}(\mathbf{r'},\mathbf{r},\omega). \label{befY}
\end{eqnarray}
Equations (\ref{fE}) and (\ref{feq2}) show that
\begin{equation} \label{fAfE}
\rmi\omega\mathbf{f}^\lambda_{\mathrm{A}}(\mathbf{r'},\mathbf{r},\omega)=\mathbf{f}^\lambda_{\mathrm{E}}(\mathbf{r'},\mathbf{r},\omega)+\frac{1}{\varepsilon_0}\left[\int_0^\infty \rmd\omega'\,\alpha(\mathbf{r'},\omega')\mathbf{f}^\lambda_{\mathrm{X}}(\mathbf{r'},\mathbf{r},\omega', \omega)\right]_\mathrm{L},
\end{equation}
which implies
\begin{equation} \label{curlfA}
\rmi\omega\nabla'\times\mathbf{f}^\lambda_{\mathrm{A}}(\mathbf{r'},\mathbf{r},\omega)=\nabla'\times\mathbf{f}^\lambda_{\mathrm{E}}(\mathbf{r'},\mathbf{r},\omega).
\end{equation}
Insertion of (\ref{alfX}), (\ref{befY}) and (\ref{fAfE}) in (\ref{fAeq}) and application of (\ref{curlfA}) produces
\begin{eqnarray}
\fl
\nabla'\times[\kappa(\mathbf{r'},\omega)\nabla'\times\mathbf{f}^\lambda_{\mathrm{E}}(\mathbf{r'},\mathbf{r},\omega)]-\frac{\omega^2}{c^2}\varepsilon(\mathbf{r'},\omega)\mathbf{f}^\lambda_{\mathrm{E}}(\mathbf{r'},\mathbf{r},\omega) \nonumber \\
=\mu_0\omega^2\alpha(\mathbf{r'},\omega)\mathbf{h}^\lambda_{\mathrm{X}}(\mathbf{r'},\mathbf{r},\omega)+\rmi \mu_0 \omega\nabla'\times[\beta(\mathbf{r'},\omega)\mathbf{h}^\lambda_{\mathrm{Y}}(\mathbf{r'},\mathbf{r},\omega)]. \label{fEeq}
\end{eqnarray}
This equation is essentially the same as the wave equation (\ref{Efreqeq})--(\ref{jfreq}) for the electric field in the frequency domain, and its solution can be written as in (\ref{EG}) using the Green bi-tensor (\ref{green}):
\begin{eqnarray} 
\mathbf{f}^\lambda_{\mathrm{E}}(\mathbf{r'},\mathbf{r},\omega)=\mu_0\int\rmd^3\mathbf{r''}\,\mathbf{G}(\mathbf{r'},\mathbf{r''}, \omega)\cdot\mathbf{s}^\lambda(\mathbf{r''},\mathbf{r},\omega),  \label{fEG} \\
\mathbf{s}^\lambda(\mathbf{r'},\mathbf{r},\omega)=\omega^2\alpha(\mathbf{r'},\omega)\mathbf{h}^\lambda_{\mathrm{X}}(\mathbf{r'},\mathbf{r},\omega)+\rmi\omega\nabla'\times[\beta(\mathbf{r'},\omega)\mathbf{h}^\lambda_{\mathrm{Y}}(\mathbf{r'},\mathbf{r},\omega)]. \label{sdef}
\end{eqnarray}

All of the bi-tensorial coefficients can now be written in terms of $\mathbf{h}^\lambda_{\mathrm{X}}$ and $\mathbf{h}^\lambda_{\mathrm{Y}}$; to avoid overly lengthy expressions we will write all the coefficients in terms of $\mathbf{h}^\lambda_{\mathrm{X}}$, $\mathbf{h}^\lambda_{\mathrm{Y}}$ and $\mathbf{f}^\lambda_{\mathrm{E}}$, bearing in mind that $\mathbf{f}^\lambda_{\mathrm{E}}$ is itself determined by $\mathbf{h}^\lambda_{\mathrm{X}}$ and $\mathbf{h}^\lambda_{\mathrm{Y}}$ through (\ref{fEG})--(\ref{sdef}). The right-hand sides of (\ref{feq1})--(\ref{feq6}) are written in terms of $\mathbf{h}^\lambda_{\mathrm{X}}$, $\mathbf{h}^\lambda_{\mathrm{Y}}$ and $\mathbf{f}^\lambda_{\mathrm{E}}$ using (\ref{fXsol}), (\ref{fYsol}), (\ref{alfX}), (\ref{befY}), (\ref{curlfA}), (\ref{fEeq}), with the results
\begin{eqnarray}
\fl
\mathbf{f}^{\lambda}_{\mathrm{\Pi_A}}(\mathbf{r'},\mathbf{r},\omega)=-\varepsilon_0\varepsilon(\mathbf{r'},\omega)\mathbf{f}^\lambda_{\mathrm{E}}(\mathbf{r'},\mathbf{r},\omega)-\alpha(\mathbf{r'},\omega)\mathbf{h}^\lambda_{\mathrm{X}}(\mathbf{r'},\mathbf{r},\omega),  \label{fPAf} \\
\fl
\mathbf{f}^\lambda_{\mathrm{A}}(\mathbf{r'},\mathbf{r},\omega)=-\frac{\rmi}{\omega}[\mathbf{f}^\lambda_{\mathrm{E}}(\mathbf{r'},\mathbf{r},\omega)]_\mathrm{T}, \label{fAf} \\
\fl
\mathbf{f}^{\lambda}_{\mathrm{\Pi_X}}(\mathbf{r'},\mathbf{r},\omega',\omega)=\frac{\rmi}{\omega}\alpha(\mathbf{r'},\omega')\mathbf{f}^\lambda_{\mathrm{E}}(\mathbf{r'},\mathbf{r},\omega)-\rmi \omega\mathbf{h}^\lambda_{\mathrm{X}}(\mathbf{r'},\mathbf{r},\omega)\delta(\omega-\omega') \nonumber \\[3pt]
-\frac{\rmi \omega'\alpha(\mathbf{r'},\omega')}{2\omega}\left[\frac{1}{\omega'-\omega-\rmi \omega0^+}+\frac{1}{\omega'+\omega-\rmi \omega0^+}\right]\mathbf{f}^\lambda_{\mathrm{E}}(\mathbf{r'},\mathbf{r},\omega),  \label{fPXf} \\[3pt]
\fl
\mathbf{f}^\lambda_{\mathrm{X}}(\mathbf{r'},\mathbf{r},\omega', \omega)=\frac{\rmi}{\omega}\mathbf{f}^\lambda_{\mathrm{\Pi_X}}(\mathbf{r'},\mathbf{r},\omega', \omega), \label{fXf} \\[3pt]
\fl
\mathbf{f}^{\lambda}_{\mathrm{\Pi_Y}}(\mathbf{r'},\mathbf{r},\omega',\omega)=\frac{1}{\omega^2}\beta(\mathbf{r'},\omega')\nabla'\times\mathbf{f}^\lambda_{\mathrm{E}}(\mathbf{r'},\mathbf{r},\omega)-\rmi \omega\mathbf{h}^\lambda_{\mathrm{Y}}(\mathbf{r'},\mathbf{r},\omega)\delta(\omega-\omega') \nonumber \\[3pt]
-\frac{\omega'\beta(\mathbf{r'},\omega')}{2\omega^2}\left[\frac{1}{\omega'-\omega-\rmi \omega0^+}+\frac{1}{\omega'+\omega-\rmi \omega0^+}\right]\nabla'\times\mathbf{f}^\lambda_{\mathrm{E}}(\mathbf{r'},\mathbf{r},\omega),  \label{fPYf} \\[3pt]
\fl
\mathbf{f}^\lambda_{\mathrm{Y}}(\mathbf{r'},\mathbf{r},\omega', \omega)=\frac{\rmi}{\omega}\mathbf{f}^\lambda_{\mathrm{\Pi_Y}}(\mathbf{r'},\mathbf{r},\omega', \omega). \label{fYf}
\end{eqnarray}
Equation (\ref{fEeq}) shows that the right-hand side of (\ref{fPAf}) is transverse, a fact that is used in deriving (\ref{fPAf}) and (\ref{fAf}). The pole prescription in (\ref{fXsol}) and (\ref{fYsol}) has been rewritten in (\ref{fPXf}) and (\ref{fPYf}) as shown in the first equality in (\ref{Ps}).

All that remains is to deduce the bi-tensors $\mathbf{h}^\lambda_{\mathrm{X}}$ and $\mathbf{h}^\lambda_{\mathrm{Y}}$; once these have been specified the bi-tensorial coefficients are found from (\ref{fEG})--(\ref{fYf}). The conditions that determine the form of  $\mathbf{h}^\lambda_{\mathrm{X}}$ and $\mathbf{h}^\lambda_{\mathrm{Y}}$ follow from substituting the expansion (\ref{Ccan}) and its hermitian conjugate into the commutation relations (\ref{CC}). When this substitution is made in the first of (\ref{CC}), and the relations (\ref{fPAf}), (\ref{fAf}), (\ref{fXf}) and (\ref{fYf}) are used, we obtain the requirement
\begin{eqnarray}
\fl
\int\rmd^3\mathbf{r''}&\left\{-\frac{\rmi}{\omega'}\left[\varepsilon_0\varepsilon^*(\mathbf{r''},\omega)f^{\lambda*}_{\mathrm{E}ki}(\mathbf{r''},\mathbf{r},\omega)+\alpha(\mathbf{r''},\omega)h^{\lambda*}_{\mathrm{X}ki}(\mathbf{r''},\mathbf{r},\omega)\right]f^{\lambda'}_{\mathrm{E}kj}(\mathbf{r''},\mathbf{r'},\omega')  \right. \nonumber \\[3pt]
\fl
&-\frac{\rmi}{\omega}f^{\lambda*}_{\mathrm{E}ki}(\mathbf{r''},\mathbf{r},\omega)\left[\varepsilon_0\varepsilon(\mathbf{r''},\omega')f^{\lambda'}_{\mathrm{E}kj}(\mathbf{r''},\mathbf{r'},\omega')+\alpha(\mathbf{r''},\omega')h^{\lambda'}_{\mathrm{X}kj}(\mathbf{r''},\mathbf{r'},\omega')\right]  \nonumber \\[3pt]
\fl
&-\left(\frac{\rmi}{\omega'}+\frac{\rmi}{\omega}\right)\int_0^\infty\rmd\omega''\left[f^{\lambda*}_{\mathrm{\Pi_X}ki}(\mathbf{r''},\mathbf{r},\omega'', \omega)f^{\lambda'}_{\mathrm{\Pi_X}kj}(\mathbf{r''},\mathbf{r'},\omega'', \omega') \right. \nonumber \\[3pt]
\fl&\qquad\qquad \qquad \qquad \quad\ \  +\left.f^{\lambda*}_{\mathrm{\Pi_Y}ki}(\mathbf{r''},\mathbf{r},\omega'', \omega)f^{\lambda'}_{\mathrm{\Pi_Y}kj}(\mathbf{r''},\mathbf{r'},\omega'', \omega') \right] {\Bigg\}} \nonumber \\[3pt]
\fl
&=-\rmi\hbar\delta_{ij}\delta_{\lambda\lambda'} \delta(\omega-\omega') \delta(\mathbf{r}-\mathbf{r'}).  \label{CCcon1}
\end{eqnarray}
The quantities in square brackets in the first two lines of (\ref{CCcon1}) are transverse because of (\ref{fEeq}); they therefore project out the transverse parts of the bi-tensors $\mathbf{f}^{\lambda'}_{\mathrm{E}}$ and  $\mathbf{f}^{\lambda*}_{\mathrm{E}}$ with which they are combined in a scalar product. For this reason the transverse restriction on  $\mathbf{f}^{\lambda'}_{\mathrm{E}}$ and  $\mathbf{f}^{\lambda*}_{\mathrm{E}}$, which was present because of (\ref{fAf}), has been removed in (\ref{CCcon1}). To find the condition on $\mathbf{h}^\lambda_{\mathrm{X}}$ and $\mathbf{h}^\lambda_{\mathrm{Y}}$ that results from (\ref{CCcon1}), we must insert (\ref{fPXf}) and (\ref{fPYf}) and perform a simplification that makes use of (\ref{fEeq}); the details of this tedious calculation are described in~\ref{ap:A} and the result is
\begin{eqnarray}
\fl
\int\rmd^3\mathbf{r''}\left[h^{\lambda*}_{\mathrm{X}ki}(\mathbf{r''},\mathbf{r},\omega)h^{\lambda'}_{\mathrm{X}kj}(\mathbf{r''},\mathbf{r'},\omega)+h^{\lambda*}_{\mathrm{Y}ki}(\mathbf{r''},\mathbf{r},\omega)h^{\lambda'}_{\mathrm{Y}kj}(\mathbf{r''},\mathbf{r'},\omega)\right]  \nonumber \\[3pt]
=\frac{\hbar}{2 \omega}\delta_{ij}\delta_{\lambda\lambda'} \delta(\mathbf{r}-\mathbf{r'}).  \label{hcon}
\end{eqnarray}
We must also impose the condition arising from the substitution of (\ref{Ccan}) into the second commutation relation in (\ref{CC}). But this condition is found to be automatically satisfied when (\ref{fPAf})--(\ref{fYf}) and (\ref{fEeq}) are utilized, and we relegate the details of this calculation to~\ref{ap:B}.

The diagonalization (\ref{Hdiag}) is achieved once we find bi-tensors $\mathbf{h}^\lambda_{\mathrm{X}}$ and $\mathbf{h}^\lambda_{\mathrm{Y}}$ that satisfy (\ref{hcon}). The diagonalization operators are of course not unique since, for example, a unitary transformation $\hat{C}_{\lambda j}(\mathbf{r},\omega)\rightarrow U_{ij}\hat{C}_{\lambda j}(\mathbf{r},\omega)$,  $U^\dagger_{ik}U_{kj}=\delta_{ij}$ does not change the form (\ref{Hdiag}), and similarly a unitary transformation can be applied to $\mathbf{h}^\lambda_{\mathrm{X}}$ and $\mathbf{h}^\lambda_{\mathrm{Y}}$ without affecting the condition (\ref{hcon}). A suitable solution of (\ref{hcon}) is
\begin{equation}  \label{hsol}
\fl
h^{\lambda}_{\mathrm{X}ij}(\mathbf{r'},\mathbf{r},\omega)=\left(\frac{\hbar}{2 \omega}\right)^{1/2}\delta_{\lambda\mathrm{e}}\delta_{ij} \delta(\mathbf{r}-\mathbf{r'}), \quad\ 
h^{\lambda}_{\mathrm{Y}ij}(\mathbf{r'},\mathbf{r},\omega)=\left(\frac{\hbar}{2 \omega}\right)^{1/2}\delta_{\lambda\mathrm{m}}\delta_{ij} \delta(\mathbf{r}-\mathbf{r'}),
\end{equation}
which completes the diagonalization procedure.

We can now write the electromagnetic field operators in terms of the diagonalizing operators $\mathbf{\hat{C}}_\lambda(\mathbf{r},t,\omega)$. The bi-tensor $\mathbf{s}^\lambda$ in (\ref{fEG})--(\ref{sdef}) is seen from (\ref{hsol}) to be
\begin{eqnarray}
\mathbf{s}^\mathrm{e}(\mathbf{r'},\mathbf{r},\omega)= \omega^2\left[\frac{\hbar\varepsilon_0}{\pi}\varepsilon_\mathrm{I}(\mathbf{r'},\omega)\right]^{1/2}\mathds{1}\delta(\mathbf{r}-\mathbf{r'}),   \\[3pt]
\mathbf{s}^\mathrm{m}(\mathbf{r'},\mathbf{r},\omega)= \rmi\omega\nabla'\times\left\{\left[-\frac{\hbar\kappa_0}{\pi}\kappa_\mathrm{I}(\mathbf{r'},\omega)\right]^{1/2}\mathds{1}\delta(\mathbf{r}-\mathbf{r'})\right\},
\end{eqnarray}
which determines through (\ref{fEG}) the bi-tensor $\mathbf{f}^\lambda_{\mathrm{E}}$. Having obtained $\mathbf{f}^\lambda_{\mathrm{E}}$, the electric-field version of (\ref{AC}) together with (\ref{Ct}) show that
\begin{eqnarray}
\fl
\mathbf{\hat{E}}(\mathbf{r},t)=\frac{\mu_0}{2\pi}\int_0^\infty\rmd \omega\int\rmd^3\mathbf{r'}\left[\rmi \omega \mathbf{G}(\mathbf{r},\mathbf{r'},\omega)\cdot\mathbf{\hat{j}} (\mathbf{r'}, \omega)\exp(-\rmi \omega t)+\mbox{h.c.}\right],  \label{EopG} \\[3pt]
\fl
\mathbf{\hat{j}} (\mathbf{r}, \omega)=-2\pi\rmi \omega\left[\frac{\hbar\varepsilon_0}{\pi}\varepsilon_\mathrm{I}(\mathbf{r},\omega)\right]^{1/2}\mathbf{\hat{C}}_\mathrm{e}(\mathbf{r},\omega) 
+2\pi\nabla\times\left\{\left[-\frac{\hbar\kappa_0}{\pi}\kappa_\mathrm{I}(\mathbf{r},\omega)\right]^{1/2}\mathbf{\hat{C}}_\mathrm{m}(\mathbf{r},\omega)\right\}. \label{jopdef}
\end{eqnarray}
The current operator defined in (\ref{jopdef}) is easily seen from (\ref{CC}) to obey the commutation relations
\begin{eqnarray}
\fl
\left[\mathbf{\hat{j}} (\mathbf{r}, \omega),\mathbf{\hat{j}}^\dagger(\mathbf{r'}, \omega')\right]=&4\pi\hbar\delta(\omega-\omega')\Bigg\{\omega^2 \varepsilon_0\varepsilon_\mathrm{I}(\mathbf{r},\omega)\mathds{1}\delta(\mathbf{r}-\mathbf{r'})  \nonumber \\
&+ \left.\kappa_0\nabla\times\left[\sqrt{-\kappa_\mathrm{I}(\mathbf{r},\omega)}\,\mathds{1}\delta(\mathbf{r}-\mathbf{r'})\sqrt{-\kappa_\mathrm{I}(\mathbf{r'},\omega')}\right]\times\stackrel{\leftarrow}{\nabla'}\right\}, \label{jj1} \\
\fl
\left[\mathbf{\hat{j}} (\mathbf{r}, \omega),\mathbf{\hat{j}}(\mathbf{r'}, \omega')\right]=&0,  \label{jj2}
\end{eqnarray}
where the notation $\times\stackrel{\leftarrow}{\nabla'}$ denotes a curl with respect to the right-hand index, so that $\mathbf{V}(\mathbf{r})\times\stackrel{\leftarrow}{\nabla}= \nabla \times\mathbf{V}(\mathbf{r})$ for a vector $\mathbf{V}(\mathbf{r})$. The transverse and longitudinal parts of $\mathbf{\hat{E}}(\mathbf{r},t)$ are related to the electromagnetic potentials by (\ref{ETL}) and the magnetic field operator can be found from 
\begin{equation} \label{BE}
\mathbf{\hat{B}}(\mathbf{r},\omega)=-\rmi\nabla\times\mathbf{\hat{E}}(\mathbf{r},\omega)/\omega,
\end{equation}
which is an identity in terms of the electromagnetic potentials.

Equations (\ref{EopG}) and (\ref{jopdef}) are the quantum versions of the classical results (\ref{EG}) and (\ref{jfreq}). The diagonalizing operators $\mathbf{\hat{C}}_\mathrm{e}(\mathbf{r},\omega)$ and $\mathbf{\hat{C}}_\mathrm{m}(\mathbf{r},\omega)$ are thus seen to be closely related to the classical fields $\mathbf{Z}_\omega(\mathbf{r})$ and $\mathbf{W}_\omega(\mathbf{r})$ in Section~\ref{sec:action}. It is easy to show that (\ref{EopG}) is the solution of the quantum macroscopic Maxwell equations. From (\ref{EopG}) and (\ref{BE}), and the defining equation (\ref{green}) of the Green bi-tensor, it follows that
\begin{equation}
\nabla\times\mathbf{\hat{H}}-\partial_t\mathbf{\hat{D}}=\mathbf{\hat{j}},  \label{qamp}
\end{equation}
with use of the definitions (\ref{D}) and (\ref{Hfield}). Similarly, (\ref{EopG}), (\ref{D}) and (\ref{green}) imply
\begin{equation}
\nabla\cdot\mathbf{\hat{D}}= \hat{\sigma}, \label{qgauss} 
\end{equation}
where in the frequency domain the charge density is
\begin{equation} \label{qsigma}
\hat{\sigma} (\mathbf{r}, \omega)=\frac{1}{\rmi\omega}\nabla \cdot\mathbf{\hat{j}}(\mathbf{r}, \omega) =-2\pi\nabla\cdot\left\{\left[\frac{\hbar\varepsilon_0}{\pi}\varepsilon_\mathrm{I}(\mathbf{r},\omega)\right]^{1/2}\mathbf{\hat{C}}_\mathrm{e}(\mathbf{r},\omega)\right\},
\end{equation}
so that the quantum version of the conservation law (\ref{conservation}) holds.

In the phenomenological approach to macroscopic QED~\cite{kno01,sch08}, a quantum noise current $\mathbf{\hat{j}}(\mathbf{r}, \omega)$ obeying the commutation relations (\ref{jj1})--(\ref{jj2}) is postulated, and this current together with its associated charge density $\hat{\sigma} (\mathbf{r}, \omega)=\nabla \cdot\mathbf{\hat{j}}(\mathbf{r}, \omega)/(\rmi\omega)$ are inserted into the macroscopic Maxwell equations (\ref{qamp})--(\ref{qgauss}). This leads to quantum electromagnetic field operators that are then shown to satisfy appropriate commutation relations as a consequence of (\ref{jj1})--(\ref{jj2}). The noise current is further decomposed into bosonic operators $\mathbf{\hat{C}}_\mathrm{e}(\mathbf{r},\omega)$ and $\mathbf{\hat{C}}_\mathrm{m}(\mathbf{r},\omega)$, as in (\ref{jopdef}), and (\ref{Hdiag}) is then the required Hamiltonian for these bosonic operators. Here we have canonically quantized macroscopic electromagnetism, proved that $\mathbf{\hat{C}}_\mathrm{e}(\mathbf{r},\omega)$ and $\mathbf{\hat{C}}_\mathrm{m}(\mathbf{r},\omega)$ in (\ref{EopG})--(\ref{jopdef}) are the diagonalizing operators of the Hamiltonian, and derived the current-density commutation relations (\ref{jj1})--(\ref{jj2}).

An interesting issue raised by Huttner and Barnett~\cite{hut92} is the possibility of negative-energy eigenstates in their microscopic model if the coupling constants are too large, since this can happen in the coupling of harmonic oscillators. Huttner and Barnett~\cite{hut92} found only positive frequencies in the diagonalized Hamiltonian, without restrictions on the size of the coupling constants in their theory, in which only the transverse part of the vector potential is quantized. This conclusion was supported by verifying that operator expansions analogous to (\ref{AC})--(\ref{PXC}) can be inverted to give expansions analogous to (\ref{Ccan}), with only positive frequencies present in the expansions. Here we have similarly found that only positive frequencies are required in the diagonalized Hamiltonian (\ref{Hdiag}). The relation of these results to the case of general coupled harmonic oscillators and the occurrence of negative energies may be interesting to investigate.

\section{Conclusions}
The absorption of electromagnetic energy in media is a necessary consequence of the restriction to retarded solutions of Maxwell's equations, which leads to the Kramers-Kronig relations. This means that the canonical quantization of electromagnetism in media cannot proceed by treating only the electromagnetic fields as dynamical variables, unless restrictions are made in the frequency range of applicability. By including a reservoir in addition to the electromagnetic fields, we developed macroscopic QED using the standard canonical formalism of quantum field theory. The theory applies to any linear, inhomogeneous, magnetodielectric medium with dielectric functions satisfying the Kramers-Kronig relations. The starting point of the phenomenological approach to macroscopic QED was derived from the underlying canonical theory. Important results such as the Lifshitz theory~\cite{lif55, dzy61,LL} of vacuum and thermal electromagnetic forces can therefore be provided with a canonical foundation.

\ack
This research is supported by the Scottish Government and the Royal Society of Edinburgh.

\appendix
\section{} \label{ap:A}
Here we describe how (\ref{CCcon1}) is simplified to (\ref{hcon}). The first step is to insert (\ref{fPXf}) and (\ref{fPYf}) into the terms in (\ref{CCcon1}) that depend on $\mathbf{f}^{\lambda}_{\mathrm{\Pi_X}}$ and $\mathbf{f}^{\lambda}_{\mathrm{\Pi_Y}}$ and their complex conjugates. We must note that all frequency arguments in (\ref{CCcon1}) are positive, since they are all positive in (\ref{CC}), which leads to (\ref{CCcon1}). Consider first the term
\begin{equation}  \label{fPXfPX}
\fl
\int\rmd^3\mathbf{r''}\left\{-\left(\frac{\rmi}{\omega'}+\frac{\rmi}{\omega}\right)\int_0^\infty\rmd\omega''f^{\lambda*}_{\mathrm{\Pi_X}ki}(\mathbf{r''},\mathbf{r},\omega'', \omega)f^{\lambda'}_{\mathrm{\Pi_X}kj}(\mathbf{r''},\mathbf{r'},\omega'', \omega') \right\}
\end{equation}
in (\ref{CCcon1}). Insertion of (\ref{fPXf}) into (\ref{fPXfPX}) leads to products of the pole prescriptions in (\ref{fPXf}) with their complex conjugates; these various products can be rewritten using the following identities:
\begin{eqnarray}
\fl
\frac{1}{(\omega''-\omega+\rmi0^+)(\omega''-\omega'-\rmi0^+)}=\frac{1}{\omega-\omega'-2\rmi0^+}\left(\frac{1}{\omega''-\omega+\rmi0^+}-\frac{1}{\omega''-\omega'-\rmi0^+}\right)  \label{A:Pi} \\[3pt]
\fl
\frac{1}{(\omega''-\omega+\rmi0^+)(\omega''+\omega'-\rmi0^+)}=\frac{1}{\omega+\omega'-2\rmi0^+}\left(\frac{1}{\omega''-\omega+\rmi0^+}-\frac{1}{\omega''+\omega'-\rmi0^+}\right)  \\[3pt]
\fl
\frac{1}{(\omega''+\omega+\rmi0^+)(\omega''-\omega'-\rmi0^+)}=\frac{1}{\omega+\omega'+2\rmi0^+}\left(\frac{1}{\omega''-\omega'-\rmi0^+}-\frac{1}{\omega''+\omega+\rmi0^+}\right)  \\[3pt]
\fl
\frac{1}{(\omega''+\omega+\rmi0^+)(\omega''+\omega'-\rmi0^+)}=\frac{1}{\omega-\omega'+2\rmi0^+}\left(\frac{1}{\omega''+\omega'-\rmi0^+}-\frac{1}{\omega''+\omega+\rmi0^+}\right)   \label{A:Pf}
\end{eqnarray}
By expanding the quantities in brackets in these formulae in terms of principle values and delta functions, as in (\ref{Ps}), and similarly expanding the other pole prescriptions in (\ref{fPXfPX}) that involve $\omega''$, the integration over $\omega''$ can be evaluated. As all of the frequencies $\omega$, $\omega'$ and $\omega''$ are positive, any delta function whose argument is a sum of two frequencies does not contribute. Making use of the Kramers-Kronig relation (\ref{KK}) in performing the $\omega''$ integration, and after straightforward but lengthy manipulations, we obtain for (\ref{fPXfPX})
\begin{eqnarray}
\fl
\int\rmd^3\mathbf{r''}\left\{-\left(\frac{\rmi}{\omega'}+\frac{\rmi}{\omega}\right)\int_0^\infty\rmd\omega''f^{\lambda*}_{\mathrm{\Pi_X}ki}(\mathbf{r''},\mathbf{r},\omega'', \omega)f^{\lambda'}_{\mathrm{\Pi_X}kj}(\mathbf{r''},\mathbf{r'},\omega'', \omega') \right\}  \nonumber  \\[3pt]
\fl
=-\rmi\int\rmd^3\mathbf{r''}\left\{\left[\mathrm{P}\frac{1}{\omega-\omega'}+\rmi\pi\delta(\omega-\omega')\right] \varepsilon_0\left[\varepsilon^*(\mathbf{r''}, \omega)-\varepsilon(\mathbf{r''}, \omega')\right]\right.     \nonumber  \\[3pt]
\times f^{\lambda*}_{\mathrm{E}ki}(\mathbf{r''},\mathbf{r},\omega)f^{\lambda'}_{\mathrm{E}kj}(\mathbf{r''},\mathbf{r'},\omega')   \nonumber  \\
-\left[\mathrm{P}\frac{1}{\omega-\omega'}+\rmi\pi\delta(\omega-\omega')\right]\alpha (\mathbf{r''}, \omega')f^{\lambda*}_{\mathrm{E}ki}(\mathbf{r''},\mathbf{r},\omega)h^{\lambda'}_{\mathrm{X}kj}(\mathbf{r''},\mathbf{r'},\omega') \nonumber  \\[3pt]
+\left[\mathrm{P}\frac{1}{\omega-\omega'}+\rmi\pi\delta(\omega-\omega')\right]\alpha (\mathbf{r''}, \omega)h^{\lambda*}_{\mathrm{X}ki}(\mathbf{r''},\mathbf{r},\omega)f^{\lambda'}_{\mathrm{E}kj}(\mathbf{r''},\mathbf{r'},\omega')   \nonumber \\
+2\omega\delta(\omega-\omega')h^{\lambda*}_{\mathrm{X}ki}(\mathbf{r''},\mathbf{r},\omega)h^{\lambda'}_{\mathrm{X}kj}(\mathbf{r''},\mathbf{r'},\omega')\Bigg\}.  \label{fPXfPXsim}
\end{eqnarray} 
Proceeding similarly with the $\mathbf{f}^{\lambda}_{\mathrm{\Pi_Y}}$-dependent term in (\ref{CCcon1}) yields
\begin{eqnarray}
\fl
\int\rmd^3\mathbf{r''}\left\{-\left(\frac{\rmi}{\omega'}+\frac{\rmi}{\omega}\right)\int_0^\infty\rmd\omega''f^{\lambda*}_{\mathrm{\Pi_Y}ki}(\mathbf{r''},\mathbf{r},\omega'', \omega)f^{\lambda'}_{\mathrm{\Pi_Y}kj}(\mathbf{r''},\mathbf{r'},\omega'', \omega') \right\}  \nonumber  \\[3pt]
\fl
=-\rmi\int\rmd^3\mathbf{r''}\left\{\frac{1}{\omega \omega'}\left[\mathrm{P}\frac{1}{\omega-\omega'}+\rmi\pi\delta(\omega-\omega')\right] \kappa_0\left[\kappa(\mathbf{r''}, \omega')-\kappa^*(\mathbf{r''}, \omega)\right]\right.     \nonumber  \\[3pt]
\times (\nabla''\times\mathbf{f}^{\lambda*}_{\mathrm{E}})_{ki}(\mathbf{r''},\mathbf{r},\omega) (\nabla''\times\mathbf{f}^{\lambda'}_{\mathrm{E}})_{kj}(\mathbf{r''},\mathbf{r'},\omega')   \nonumber  \\
-\frac{\rmi}{\omega}\left[\mathrm{P}\frac{1}{\omega-\omega'}+\rmi\pi\delta(\omega-\omega')\right]\beta (\mathbf{r''}, \omega') (\nabla''\times\mathbf{f}^{\lambda*}_{\mathrm{E}})_{ki}(\mathbf{r''},\mathbf{r},\omega)h^{\lambda'}_{\mathrm{Y}kj}(\mathbf{r''},\mathbf{r'},\omega') \nonumber  \\[3pt]
-\frac{\rmi}{\omega'}\left[\mathrm{P}\frac{1}{\omega-\omega'}+\rmi\pi\delta(\omega-\omega')\right]\beta (\mathbf{r''}, \omega)h^{\lambda*}_{\mathrm{Y}ki}(\mathbf{r''},\mathbf{r},\omega) (\nabla''\times\mathbf{f}^{\lambda'}_{\mathrm{E}})_{kj}(\mathbf{r''},\mathbf{r'},\omega')   \nonumber \\
+2\omega\delta(\omega-\omega')h^{\lambda*}_{\mathrm{Y}ki}(\mathbf{r''},\mathbf{r},\omega)h^{\lambda'}_{\mathrm{Y}kj}(\mathbf{r''},\mathbf{r'},\omega')\Bigg\}.  \label{fPYfPYsim}
\end{eqnarray} 
The results (\ref{fPXfPXsim}) and (\ref{fPYfPYsim}) must now be used in (\ref{CCcon1}). Combining the terms in the first two lines of (\ref{CCcon1}) with (\ref{fPXfPXsim}) produces the quantity
\begin{eqnarray}
\fl
-\rmi\int\rmd^3\mathbf{r''}\left\{\varepsilon_0\left[\mathrm{P}\frac{\omega}{\omega'(\omega-\omega')}\varepsilon^*(\mathbf{r''}, \omega)-\mathrm{P}\frac{\omega'}{\omega(\omega-\omega')}\varepsilon(\mathbf{r''}, \omega')  \right.\right.   \nonumber  \\[3pt]
\ \ \ +\rmi\pi[\varepsilon^*(\mathbf{r''}, \omega)-\varepsilon(\mathbf{r''}, \omega')]\delta(\omega-\omega')\Bigg]f^{\lambda*}_{\mathrm{E}ki}(\mathbf{r''},\mathbf{r},\omega)f^{\lambda'}_{\mathrm{E}kj}(\mathbf{r''},\mathbf{r'},\omega')   \nonumber  \\
-\left[\mathrm{P}\frac{\omega'}{\omega(\omega-\omega')}+\rmi\pi\delta(\omega-\omega')\right]\alpha (\mathbf{r''}, \omega')f^{\lambda*}_{\mathrm{E}ki}(\mathbf{r''},\mathbf{r},\omega)h^{\lambda'}_{\mathrm{X}kj}(\mathbf{r''},\mathbf{r'},\omega') \nonumber  \\[3pt]
+\left[\mathrm{P}\frac{\omega}{\omega'(\omega-\omega')}+\rmi\pi\delta(\omega-\omega')\right]\alpha (\mathbf{r''}, \omega)h^{\lambda*}_{\mathrm{X}ki}(\mathbf{r''},\mathbf{r},\omega)f^{\lambda'}_{\mathrm{E}kj}(\mathbf{r''},\mathbf{r'},\omega')   \nonumber \\
+2\omega\delta(\omega-\omega')h^{\lambda*}_{\mathrm{X}ki}(\mathbf{r''},\mathbf{r},\omega)h^{\lambda'}_{\mathrm{X}kj}(\mathbf{r''},\mathbf{r'},\omega')\Bigg\}.  \label{Ares}
\end{eqnarray} 
We must now add (\ref{Ares}) to (\ref{fPYfPYsim}) and thereby obtain the left-hand side of (\ref{CCcon1}). Before performing this addition we make an integration by parts in each of the terms in (\ref{fPYfPYsim}) containing a curl. It is then straightforward to see that because of (\ref{fEeq}) most terms in the sum of (\ref{Ares}) and (\ref{fPYfPYsim}) cancel. The only terms that survive are the last term in (\ref{Ares}) and the last term in (\ref{fPYfPYsim}). We thus obtain (\ref{hcon}).

\section{} \label{ap:B}
Here we describe the result of substituting (\ref{Ccan}) into the second commutation relation in (\ref{CC}). This calculation is of course very similar to the substitution of (\ref{Ccan}) into the first commutation relation in (\ref{CC}), which simplified to the condition (\ref{hcon}).  In this case, instead of the condition (\ref{CCcon1}) on the bi-tensorial coefficients, we obtain
\begin{eqnarray}
\fl
\int\rmd^3\mathbf{r''}&\left\{-\frac{\rmi}{\omega'}\left[\varepsilon_0\varepsilon^*(\mathbf{r''},\omega)f^{\lambda*}_{\mathrm{E}ki}(\mathbf{r''},\mathbf{r},\omega)+\alpha(\mathbf{r''},\omega)h^{\lambda*}_{\mathrm{X}ki}(\mathbf{r''},\mathbf{r},\omega)\right]f^{\lambda'*}_{\mathrm{E}kj}(\mathbf{r''},\mathbf{r'},\omega')  \right. \nonumber \\[3pt]
\fl
&+\frac{\rmi}{\omega}f^{\lambda*}_{\mathrm{E}ki}(\mathbf{r''},\mathbf{r},\omega)\left[\varepsilon_0\varepsilon^*(\mathbf{r''},\omega')f^{\lambda'*}_{\mathrm{E}kj}(\mathbf{r''},\mathbf{r'},\omega')+\alpha(\mathbf{r''},\omega')h^{\lambda'*}_{\mathrm{X}kj}(\mathbf{r''},\mathbf{r'},\omega')\right]  \nonumber \\[3pt]
\fl
&-\left(\frac{\rmi}{\omega'}-\frac{\rmi}{\omega}\right)\int_0^\infty\rmd\omega''\left[f^{\lambda*}_{\mathrm{\Pi_X}ki}(\mathbf{r''},\mathbf{r},\omega'', \omega)f^{\lambda'*}_{\mathrm{\Pi_X}kj}(\mathbf{r''},\mathbf{r'},\omega'', \omega') \right. \nonumber \\[3pt]
\fl&\qquad\qquad \qquad \qquad \quad\ \  +\left.f^{\lambda*}_{\mathrm{\Pi_Y}ki}(\mathbf{r''},\mathbf{r},\omega'', \omega)f^{\lambda'*}_{\mathrm{\Pi_Y}kj}(\mathbf{r''},\mathbf{r'},\omega'', \omega') \right] {\Bigg\}}=0. \label{CCcon2}
\end{eqnarray}
Proceeding as in~\ref{ap:A}, we substitute (\ref{fPXf}) and (\ref{fPYf}) into (\ref{CCcon2}). Consider first the term
\begin{equation}  \label{fPXfPX2}
\fl
\int\rmd^3\mathbf{r''}\left\{-\left(\frac{\rmi}{\omega'}-\frac{\rmi}{\omega}\right)\int_0^\infty\rmd\omega''f^{\lambda*}_{\mathrm{\Pi_X}ki}(\mathbf{r''},\mathbf{r},\omega'', \omega)f^{\lambda'*}_{\mathrm{\Pi_X}kj}(\mathbf{r''},\mathbf{r'},\omega'', \omega') \right\}.
\end{equation}
Insertion of (\ref{fPXf}) into (\ref{fPXfPX2}) leads to products of the pole prescriptions  that can be rewritten using the following identities:
\begin{eqnarray}
\fl
\frac{1}{(\omega''-\omega+\rmi0^+)(\omega''-\omega'+\rmi0^+)}=\frac{1}{\omega-\omega'}\left(\frac{1}{\omega''-\omega+\rmi0^+}-\frac{1}{\omega''-\omega'+\rmi0^+}\right)  \label{B:P1} \\[3pt]
\fl
\frac{1}{(\omega''-\omega+\rmi0^+)(\omega''+\omega'+\rmi0^+)}=\frac{1}{\omega+\omega'}\left(\frac{1}{\omega''-\omega+\rmi0^+}-\frac{1}{\omega''+\omega'+\rmi0^+}\right)   \label{B:P2}\\[3pt]
\fl
\frac{1}{(\omega''+\omega+\rmi0^+)(\omega''-\omega'+\rmi0^+)}=\frac{1}{\omega+\omega'}\left(\frac{1}{\omega''-\omega'+\rmi0^+}-\frac{1}{\omega''+\omega+\rmi0^+}\right)   \label{B:P3} \\[3pt]
\fl
\frac{1}{(\omega''+\omega+\rmi0^+)(\omega''+\omega'+\rmi0^+)}=\frac{1}{\omega-\omega'}\left(\frac{1}{\omega''+\omega'+\rmi0^+}-\frac{1}{\omega''+\omega+\rmi0^+}\right)   \label{B:P4}
\end{eqnarray}
In the terms $1/(\omega -\omega')$ on the right-hand sides of identities (\ref{B:P1}) and (\ref{B:P4}) there is no prescription for dealing with the pole at $\omega=\omega'$; this is not a problem however, because when (\ref{B:P1})--(\ref{B:P4}) are used in (\ref{fPXfPX2}) these factors of $1/(\omega -\omega')$ get combined with other terms in such a manner that they do not give rise to a pole. Bearing in mind that all frequencies are positive we obtain for (\ref{fPXfPX2})
\begin{eqnarray}
\fl
\int\rmd^3\mathbf{r''}\left\{-\left(\frac{\rmi}{\omega'}-\frac{\rmi}{\omega}\right)\int_0^\infty\rmd\omega''f^{\lambda*}_{\mathrm{\Pi_X}ki}(\mathbf{r''},\mathbf{r},\omega'', \omega)f^{\lambda'*}_{\mathrm{\Pi_X}kj}(\mathbf{r''},\mathbf{r'},\omega'', \omega') \right\} \nonumber  \\[3pt]
\fl
=-\rmi\int\rmd^3\mathbf{r''}\left\{-\frac{1}{\omega+\omega'} \varepsilon_0\left[\varepsilon^*(\mathbf{r''}, \omega)-\varepsilon^*(\mathbf{r''}, \omega')\right]
 f^{\lambda*}_{\mathrm{E}ki}(\mathbf{r''},\mathbf{r},\omega)f^{\lambda'*}_{\mathrm{E}kj}(\mathbf{r''},\mathbf{r'},\omega')\right.   \nonumber  \\
+\frac{1}{\omega+\omega'}\alpha (\mathbf{r''}, \omega')f^{\lambda*}_{\mathrm{E}ki}(\mathbf{r''},\mathbf{r},\omega)h^{\lambda'*}_{\mathrm{X}kj}(\mathbf{r''},\mathbf{r'},\omega') \nonumber  \\[3pt]
-\frac{1}{\omega+\omega'}\alpha (\mathbf{r''}, \omega)h^{\lambda*}_{\mathrm{X}ki}(\mathbf{r''},\mathbf{r},\omega)f^{\lambda'*}_{\mathrm{E}kj}(\mathbf{r''},\mathbf{r'},\omega') \Bigg\}.  \label{fPXfPX2sim}
\end{eqnarray} 
Note that there is no term quadratic in $\mathbf{h}^{\lambda}_{\mathrm{X}}$ in (\ref{fPXfPX2sim}), in contrast to the analogous result (\ref{fPXfPXsim}). This is because in the case of (\ref{fPXfPX2sim}) the  term quadratic in $\mathbf{h}^{\lambda}_{\mathrm{X}}$ appeared with a factor $(\omega-\omega')\delta(\omega-\omega')=0$.
The $\mathbf{f}^{\lambda}_{\mathrm{\Pi_Y}}$-dependent term in (\ref{CCcon2}) similarly yields
\begin{eqnarray}
\fl
\int\rmd^3\mathbf{r''}\left\{-\left(\frac{\rmi}{\omega'}-\frac{\rmi}{\omega}\right)\int_0^\infty\rmd\omega''f^{\lambda*}_{\mathrm{\Pi_Y}ki}(\mathbf{r''},\mathbf{r},\omega'', \omega)f^{\lambda'*}_{\mathrm{\Pi_Y}kj}(\mathbf{r''},\mathbf{r'},\omega'', \omega') \right\} \nonumber  \\[3pt]
\fl
=-\rmi\int\rmd^3\mathbf{r''}\left\{\frac{1}{\omega \omega'(\omega+\omega')} \kappa_0\left[\kappa^*(\mathbf{r''}, \omega')-\kappa^*(\mathbf{r''}, \omega)\right]\right. \nonumber  \\[3pt]
\times (\nabla''\times\mathbf{f}^{\lambda*}_{\mathrm{E}})_{ki}(\mathbf{r''},\mathbf{r},\omega) (\nabla''\times\mathbf{f}^{\lambda'*}_{\mathrm{E}})_{kj}(\mathbf{r''},\mathbf{r'},\omega')   \nonumber  \\[3pt]
+\frac{\rmi}{\omega(\omega+\omega')}\beta(\mathbf{r''}, \omega') (\nabla''\times\mathbf{f}^{\lambda*}_{\mathrm{E}})_{ki}(\mathbf{r''},\mathbf{r},\omega)h^{\lambda'*}_{\mathrm{Y}kj}(\mathbf{r''},\mathbf{r'},\omega') \nonumber  \\[3pt]
-\frac{\rmi}{\omega'(\omega+\omega')}\beta(\mathbf{r''}, \omega)h^{\lambda*}_{\mathrm{Y}ki}(\mathbf{r''},\mathbf{r},\omega) (\nabla''\times\mathbf{f}^{\lambda'*}_{\mathrm{E}})_{kj}(\mathbf{r''},\mathbf{r'},\omega') \Bigg\}.  \label{fPYfPY2sim}
\end{eqnarray} 
Combining the terms in the first two lines of (\ref{CCcon2}) with (\ref{fPXfPX2sim}) produces the quantity
\begin{eqnarray}
\fl
-\rmi\int\rmd^3\mathbf{r''}\left\{\varepsilon_0\left[\frac{\omega}{\omega'(\omega+\omega')}\varepsilon^*(\mathbf{r''}, \omega)-\frac{\omega'}{\omega(\omega+\omega')}\varepsilon^*(\mathbf{r''}, \omega')\right]f^{\lambda*}_{\mathrm{E}ki}(\mathbf{r''},\mathbf{r},\omega)f^{\lambda'*}_{\mathrm{E}kj}(\mathbf{r''},\mathbf{r'},\omega')   \right.   \nonumber  \\[3pt]
-\frac{\omega'}{\omega(\omega+\omega')}\alpha (\mathbf{r''}, \omega')f^{\lambda*}_{\mathrm{E}ki}(\mathbf{r''},\mathbf{r},\omega)h^{\lambda'*}_{\mathrm{X}kj}(\mathbf{r''},\mathbf{r'},\omega') \nonumber  \\[3pt]
+\frac{\omega}{\omega'(\omega+\omega')}\alpha (\mathbf{r''}, \omega)h^{\lambda*}_{\mathrm{X}ki}(\mathbf{r''},\mathbf{r},\omega)f^{\lambda'*}_{\mathrm{E}kj}(\mathbf{r''},\mathbf{r'},\omega')\Bigg\}.  \label{Ares2}
\end{eqnarray} 
We now add (\ref{Ares2}) to (\ref{fPYfPY2sim}) to obtain the left-hand side of (\ref{CCcon2}). An integration by parts in each of the terms in (\ref{fPYfPY2sim}) containing a curl and use of (\ref{fEeq}) shows that the result of this addition is zero. The condition (\ref{CCcon2}) therefore gives no restriction on the bi-tensors $\mathbf{h}^{\lambda}_{\mathrm{X}}$ and $\mathbf{h}^{\lambda}_{\mathrm{Y}}$.

\end{document}